\documentclass[pdflatex,sn-mathphys-num]{sn-jnl}

\usepackage{graphicx}%
\usepackage{multirow}%
\usepackage{amsmath,amssymb,amsfonts}%
\usepackage{amsthm}%
\usepackage{mathrsfs}%
\usepackage[title]{appendix}%
\usepackage{xcolor}%
\usepackage{textcomp}%
\usepackage{manyfoot}%
\usepackage{booktabs}%
\usepackage{algorithm}%
\usepackage{algorithmicx}%
\usepackage{algpseudocode}%
\usepackage{listings}%
\usepackage{braket}
\usepackage{graphicx}
\usepackage{subcaption}
\usepackage{caption}
\usepackage{geometry}
\usepackage{lipsum}
\usepackage{tikz}
\usepackage{enumitem}
\usepackage{setspace}
\usepackage{framed}
\usepackage{float}
\usetikzlibrary{positioning}
\graphicspath{{SimRts/}{../png/}{../jpeg/}}

\theoremstyle{thmstyleone}%
\theoremstyle{thmstyletwo}%

\theoremstyle{thmstylethree}%

\begin{document}

\title[Article Title]{A Two-stage Optimization Method for Wide-range Single-electron Quantum Magnetic Sensing}


\author[1]{\fnm{Shiqian} \sur{Guo}}\email{sguo26@ncsu.edu}

\author*[1]{\fnm{Jianqing} \sur{Liu}}\email{jliu96@ncsu.edu}

\author[1]{\fnm{Thinh} \sur{Le}}\email{tvle2@ncsu.edu}

\author[2]{\fnm{Huaiyu} \sur{Dai}}\email{hdai@ncsu.edu}

\affil[1]{\orgdiv{Department of Computer Science}, \orgname{North Carolina State University}} \affil[2]{\orgdiv{Department of Electrical \& Computer Engineering}, \orgname{North Carolina State University}} \affil[]{\orgaddress{\street{890 Oval Drive}, \city{Raleigh}, \postcode{27659}, \state{NC}, \country{USA}}}


\abstract{Quantum magnetic sensing based on spin systems has emerged as a new paradigm for detecting ultra-weak magnetic fields with unprecedented sensitivity, revitalizing applications in navigation, geo-localization, biology, and beyond. At the heart of quantum magnetic sensing, from the protocol perspective, lies the design of optimal sensing parameters to manifest and then estimate the underlying signals of interest (SoI). Existing studies on this front mainly rely on adaptive algorithms based on black-box AI models or formula-driven principled searches. However, when the SoI spans a wide range and the quantum sensor has physical constraints, these methods may fail to converge efficiently or optimally, resulting in prolonged interrogation times and reduced sensing accuracy. In this work, we report the design of a new protocol using a two-stage optimization method. In the 1$^{st}$ Stage, a Bayesian neural network with a fixed set of sensing parameters is used to narrow the range of SoI. In the 2$^{nd}$ Stage, a federated reinforcement learning agent is designed to fine-tune the sensing parameters within a reduced search space. The proposed protocol is developed and evaluated in a challenging context of single-shot readout of an NV-center electron spin under a constrained total sensing time budget; and yet it achieves significant improvements in both accuracy and resource efficiency for wide-range D.C. magnetic field estimation compared to the state of the art.}




\maketitle

\section{Introduction}\label{sec1}

Quantum sensors harness the fundamental principles of quantum mechanics to detect physical quantities that are imperceptible to classical sensors. Among their applications, magnetic field sensing at nanoscale resolution is particularly important, with uses spanning many applications such as navigation, medical diagnostics, and cybersecurity \cite{casacio2021quantum, rondin2014magnetometry,liu2024road}. To detect weak magnetic signals, Ramsey interferometry has been long established as one of the most sensitive approaches~\cite{degen2017quantum}. In Ramsey interferometry, a qubit in the superposition state evolves with the magnetic field to acquire phase at a rate that depends on the qubit frequency $\omega$. The precise estimation of the magnetic signal hinges on the extraction of $\omega$, a central research problem in the Ramsey interferometry design. Ramsey interferometry can be implemented by different quantum species, such as superconducting qubits and spin qubits. Among different quantum magnetic sensing platforms, 
the negatively charged nitrogen-vacancy (NV) centers in diamonds has emerged as one of the most promising platforms for its  long coherence time even at room temperature. Moreover, the NV center-based magnetometer boasts a rather low energy resolution limit (ERL) $E_R \approx \hbar$, a figure of merit that indicats exceptional performance in (magnetic) field resolution, bandwidth, and the size of the sensed region~\cite{mitchell2020colloquium}. This performance arises from two key quantum properties of NV centers: spin coherence and photoluminescence. At room temperature, NV centers exhibit remarkably long spin coherence. This stability comes from the fact that the surrounding diamond carbon isotopes (e.g. $^{12}\text{C}$) have zero nuclear spin and therefore do not affect the spin, while the rigid covalent bonds of the lattice keep the NV centers with their trapped electrons relatively isolated. In addition, the fluorescence properties of the defect enable efficient optical reading of its quantum state. These advantages have propelled NV centers to the forefront of quantum sensing, with demonstrations spanning magnetic \cite{maletinsky2012robust}, electric \cite{dolde2011electric}, and temperature \cite{acosta2010temperature} sensing at resolutions down to the nanoscale.

The NV centers exhibits a triplet ground state (S=1) with $\text{C}_{3v}$ symmetry and spin sub-levels $\text{m}_s = 0$ and $\text{m}_s = \pm 1$ along the N-V axis. In the absence of external magnetic fields and at room temperature, the crystal field interaction induces a zero-field splitting of $\approx D = 2.87$ GHz between the $\text{m}_s = 0$ and degenerate $\text{m}_s = \pm 1$ spin sub-levels due to spin-spin interactions within the diamond lattice \cite{manson2006nitrogen}. Coherent manipulation of these spin states is achieved through resonant microwave radiation at GHz frequencies \cite{guo2023overview} and spin-state readout is facilitated by spin-dependent photoluminescence, as the $\text{m}_s = \pm 1$ states exhibit reduced fluorescence intensity compared to the $\text{m}_s = 0$ state due to different relaxation pathways through metastable singlet states \cite{hopper2018spin}. When a D.C. magnetic field $B$ is applied, the Zeeman effect lifts the degeneracy of the $\text{m}_s = \pm 1$ spin sub-levels, splitting them into two distinct levels. Precision magnetic field sensing with NV centers is performed by quantitatively measuring the Zeeman shifts $E_z = \hbar \gamma B = \hbar 2\pi \omega$, where $\gamma$ is the NV gyromagnetic ratio and $\omega$ is the Larmor frequency, using Ramsey interferometry \cite{bonato2016optimized}. 



The exceptional spin coherence and photoluminescent readout of NV centers create a versatile platform for quantum magnetometry. However, realizing this potential requires addressing two interrelated challenges: quantum metrology (i.e., optimizing sensing protocols) \cite{taylor2016quantum, giovannetti2011advances, toth2014quantum} and parameter estimation (i.e., inferring magnetic fields from noisy measurements) \cite{morelli2021bayesian, lee2022quantum}. Quantum metrology focuses on maximizing sensitivity by careful design of control parameters, such as the sensing time and microwave phase in pulsed sequences like Ramsey interferometry. For NV centers, the choice of sensing time needs to balance competing demands: longer sensing time increases phase accumulation (enhancing sensitivity) but risks decoherence, while shorter sensing time reduces noise susceptibility at the cost of signal strength. Recent advances in adaptive control elucidate potential pathways to overcome these limitations \cite{berritta2024real, salvia2023critical, maclellan2024end, xu2019generalizable}. Parameter estimation involves solving the inverse problem of reconstructing the D.C. magnetic field from measurement outcomes. This task is typically approached through frequentist or Bayesian frameworks. The frequentist approach leverages the statistical properties of measurement outcomes to perform estimation \cite{nolan2021frequentist}. In contrast, the adaptive Bayesian approach tailors metrological parameters based on the past knowledge of the estimate. It has been shown that such an adaptive estimation based on prior knowledge can achieve the Heisenberg limit without the need for entangled states \cite{higgins2007entanglement}. In this line of research, McMichael et al. \cite{mcmichael2022simplified} devised an efficient estimation method for quantum sensing experiments by reducing computational complexity associated with the conventional Bayesian framework. Moreover, adaptive Bayesian methods have demonstrated significant improvements in estimation accuracy within a fixed time budget and across a wide field range \cite{bonato2016optimized}. Real-time adaptive Bayesian techniques have also facilitated the estimation of decoherence timescales for single qubits \cite{arshad2024real}, while results in \cite{zohar2023real} underscore the generality of the adaptive Bayesian method for the quantum sensors without single-shot-readout.

Although adaptive Bayesian estimation offers high accuracy and efficiency, implementing Bayesian updates and designing sequential experiments analytically often entails prohibitive computational complexity. Machine learning, with its demonstrated ability to extract meaningful patterns from large and complex datasets, offers a promising solution, particularly when integrated into the post-processing of measurement data. Reinforcement learning (RL), an active learning paradigm that optimize experimental policies (i.e., sequences of actions) by interacting with dynamic environments to maximize cumulative rewards, has emerged as a powerful tool for designing quantum sensing protocols \cite{fiderer2021neural, belliardo2024model, cimini2023deep}. Strengthened by deep neural networks, deep RL can effectively approximate intricate policy functions from a large dataset and exhibits strong robustness to noise in measurement data, further enhancing its applicability to real-world quantum experiments \cite{metz2023self, fosel2018reinforcement, reuer2023realizing, fallani2022learning}. However, the limited learning capacity of current deep RL approaches restricts their use cases to the predefined and relatively narrow ranges of SoI. As a result, the RL agents proposed in \cite{fiderer2021neural, belliardo2024model} struggle to achieve optimal solutions when the sensing task involves a SoI spanning a wide range. 

To overcome above challenges, this work presents a two-stage optimization method to achieve high accuracy in D.C. magnetic field sensing under limited sensing time budget and with only an NV-center spin for single readout. The basic idea is as follows. In the $1^{\text{st}}$ Stage, we fix the control parameters according to the guessed largest possible value of the unknown field $\omega_\text{max}$, make a series of Ramsey measurements, and then use a trained Bayesian neural network (BNN) estimator to generate a rough estimate of the SoI. However, such initial estimation remains unreliable due to non-optimal control parameters, imperfections in BNN training, increased discretization errors inherent to wide estimation ranges, and other contributing factors. To overcome these limitations, our protocol introduces a $2^{\text{nd}}$ Stage consisting of an adaptive algorithm based on a federated RL agent. This stage aims to fine-tune the control parameters within a narrower range informed by the coarse estimate from the first stage, thereby achieving a more precise estimation of the SoI. A key challenge arises from the fact that the estimated subrange (defined by the estimated value and its uncertainty) from the first stage is both non-deterministic and initially unknown. Therefore, conventional RL agents, typically trained in static environments with fixed sensing ranges may suffer from bad convergence in this context. To address this challenge, we incorporate federated learning into the RL training process. Each subrange constitutes a distinct RL environment, characterized by unique optimal sensing times, reward landscapes, and measurement noise profiles. In so doing, we anticipate superior performance over standard RL methods, which train a single RL agent on aggregated data and can often lead to gradient conflicts, slower convergence, and degraded generalization when task-specific policies differ significantly.

\section{Results}\label{sec2}

\begin{figure}[h]
	\centering
	\includegraphics[width=\columnwidth]{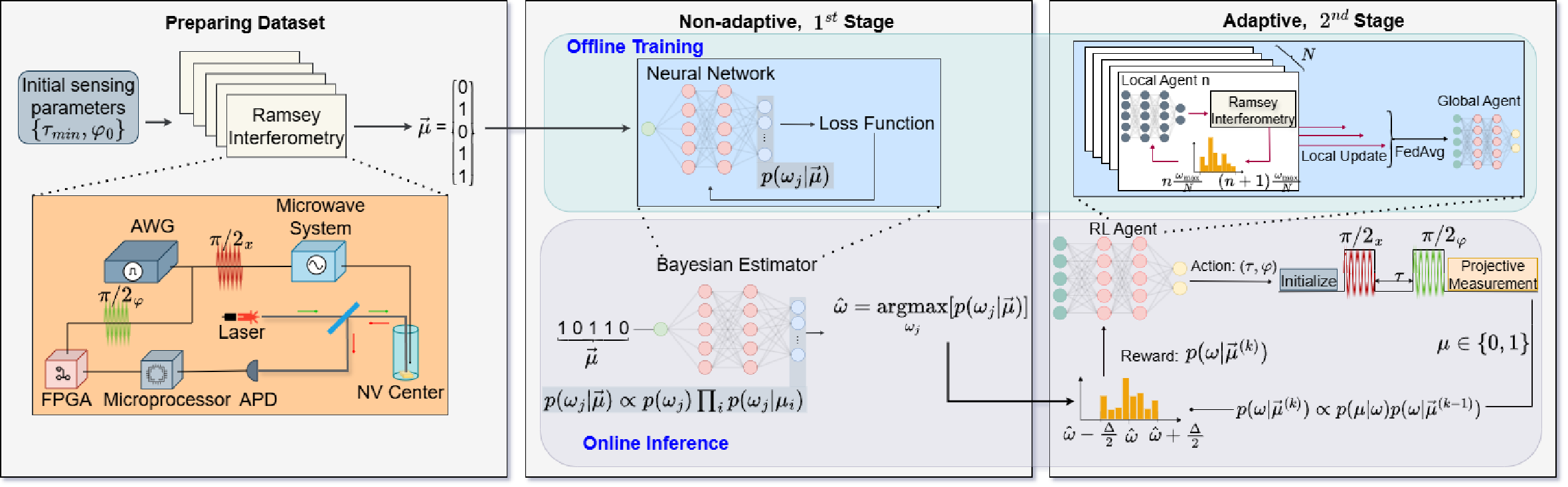}
	\caption{Framework of the proposed two-stage optimization method. The protocol consists of an offline training phase and an online inference phase. Prior to execution, two neural networks must be trained offline: one for the BNN estimator and another for the federated RL agent. To train the neural network for the BNN estimator, a large dataset is constructed using measurement outcomes from Ramsey experiments. These experiments are performed using a quantum sensor based on a single NV-center electron spin system. Each training sample is encoded as a one-hot vector representing a sequence of measurement outcomes \( \vec{\mu} \), with the corresponding label being the posterior probability distribution \( p(\omega_j \mid \vec{\mu}) \). Here, \( \omega_j \) denotes a discretized value from the magnetic field range \( (0, \omega_{\text{max}}) \). The neural network is trained to minimize a cross-entropy loss, effectively performing Bayesian inversion. For the training of the federated RL agent, the total magnetic field range is partitioned into \( N \) equal-width intervals. Each interval $[ n\omega_{\text{max}}/{N}, (n+1)\omega_{\text{max}}/{N} )$, defines a local environment for the \( n \)-th local RL agent. A dedicated RL agent is trained within each local environment, where the input after the \( k \)-th Ramsey experiment is the posterior distribution \( p(\omega \mid \vec{\mu}^{(k)}) \), approximated using a particle filter. The agent outputs an action \( (\tau, \varphi) \), representing the sensing time and control phase for the next experiment. Upon receiving a measurement outcome \( \mu \), the posterior distribution is updated using Bayes' rule. Ramsey experiment is repeated until the predefined time resources are exhausted. Each local RL agent maximizes its cumulative reward and computes a local network parameter update. The global RL agent is then updated by aggregating all local updates using the FedAvg algorithm. Once offline training is completed, the protocol proceeds to the online inference phase. In the non-adaptive 1$^{st}$ Stage, Ramsey experiments are performed using fixed sensing parameters \( \tau_{\text{min}} = {\pi}/{\omega_{\text{max}}} \) and \( \varphi_0 = 0 \) to generate a sequence of measurement outcomes \( \vec{\mu} \). This sequence is input into the trained BNN estimator to produce an initial estimate \( \hat{\omega} \), which narrows the magnetic field range to a subrange \( \left[ \hat{\omega} - \frac{\Delta}{2}, \hat{\omega} + \frac{\Delta}{2} \right] \). In the adaptive 2$^{nd}$ Stage, the trained global RL agent uses the posterior within this subrange to determine optimized sensing designs \( (\tau, \varphi) \), thereby improving the estimation accuracy through adaptive Ramsey experiments.}\label{fig:summary}
\end{figure}
\subsection{Background and State of the Art}
The proposed protocol is poised to efficiently derive the optimal sensing time and control phase of an NV-center electron spin for single-shot Ramsey measurement. Suppose that the two basis states of a spin are $\ket{1}$ and $\ket{0}$. Ramsey measurement begins with a $\frac{\pi}{2}$ pulse, preparing the NV center in a superposition state $\frac{1}{\sqrt{2}}\left(\ket{0}+\ket{1}\right)$. The Zeeman shift, defined as $\omega := \gamma B$, where $\gamma$ is the gyromagnetic ratio and $B$ is the D.C. magnetic field strength, then drives the evolution of the system. After a sensing time $\tau$, the state evolves to $\frac{1}{\sqrt{2}}\left(\ket{0}+e^{i\theta}\ket{1}\right)$, where $\theta = \omega \tau$ is the accumulated phase. Next, using a second $\frac{\pi}{2}$ pulse, the state is converted back to a measurable state $\frac{1}{2} (1 + e^{i\theta})\ket{0}+\frac{1}{2} (1-e^{i\theta})\ket{1}$. Finally, the spin is read out in an appropriate basis to extract the phase $\theta$. The probability of obtaining outcome $0$ is given by
\begin{equation}
p(0 \mid \omega) = \frac{1}{2} + \frac{1}{2} e^{-\tau / T_2} \cos(\omega \tau + \varphi),\label{equ:1}
\end{equation}
and $p(1 \mid \omega) = 1 - p(0 \mid \omega)$. $\varphi$ is the control phase applied during the final $\frac{\pi}{2}$ pulse and $T_2$ represents the spin coherence time. The outcome of the measurement $m_s = 0$ or $m_s = 1$ is probabilistic and follows a periodic function weighted by a decay coefficient related to spin decoherence. To estimate the SoI $\omega$, one has to repeat Ramsey measurements for many rounds with possibly different configurations of $\{\tau, \varphi\}$, and then analyze the outcome statistics for inference. The research problem is now boiled down to two intertwined tasks, with one being the optimization of sensing parameters $\{\tau, \varphi\}$ and the other being the parameter estimation of $\omega$. In this work, we use estimation accuracy as the performance metric, captured by the mean square error (MSE) that is defined as $\text{MSE}(\hat{\omega}, \omega) :=\mathbb{E}[|\hat{\omega} - \omega|^2] $. Here, $\hat{\omega}$ is the estimated value, while $\omega$ (or equivalently $B$) is the true value of the D.C. magnetic field.

The existing literature has presented several solutions to this research problem, particularly in the context of quantum magnetic sensing \cite{bonato2016optimized,fiderer2021neural,belliardo2024model}. Before reporting the results of this work, we first present three protocols whose underlying algorithm ideas reflect the mainstream that are broadly adopted in the community. The first one is the adaptive approach, commonly implemented as a closed loop that feeds back loss/reward for iterative update of control parameters. This is exemplified by the work by Bonato et al. \cite{bonato2016optimized}. They utilized the particle swarm optimization method that updated $\varphi$ by a fixed step size per iteration while selecting $\tau$ from a predetermined set. The stopping criteria of their algorithm was dictated by a closed-form equation borrowed from an established knowledge in \cite{cappellaro2012spin}. Another algorithm bearing the similar adaptive nature is the use of RL, exemplified by Belliardo et al. in \cite{belliardo2024model} (and an earlier work by Fiderer et al. \cite{fiderer2021neural} in a general quantum metrology context). Specifically, Belliardo et al. utilized a vanilla RL network to optimize both the sensing time $\tau$ and the controlled phase $\varphi$ \cite{belliardo2024model}. This algorithm follows a black-box model, and its convergence depends on the loss function and the convergence threshold. In addition to the adaptive approach, another notable line of method is from the work by Nolan et al. \cite{nolan2021frequentist} (though it is not directly applied to quantum magnetic sensing), coined as ``NN-shots'' protocol. It belongs to a non-adaptive approach in which a BNN is trained to perform parameter estimation, with the sensing time fixed at $\tau = \pi/\omega_{\text{max}}$. Compared with the conventional Bayesian updates that typically require an impractically large amount of observation data, the BNN in \cite{nolan2021frequentist} can be trained using individual measurement outcomes without relying on an explicit physical model or fitting function. This model-agnostic feature makes the BNN particularly well suited for estimation tasks in scenarios involving finite detection resolution and noisy probe states. In this work, we will use the works by Bonato et al. \cite{bonato2016optimized}, Belliardo et al. in \cite{belliardo2024model}, and NN-shots by Nolan et al. \cite{nolan2021frequentist}, representing different levels of algorithmic adaptiveness, as the baselines for performance comparison. 
\subsection{Summary of the Proposed Protocol}
An overview of this work's protocol is shown in Fig. \ref{fig:summary}. In summary, it is based on a two-stage progressive optimization method. Each stage consists of an offline training phase and an online inference phase, as shown on the green panel and gray panel, respectively, in Fig. \ref{fig:summary}. For presentation clarity, we focus on illustrating the runtime procedure of this protocol while directing the readers to Supplementary Section 1 and 2 of the Supplemental document for the details of dataset preparation and model training. To begin with, we run a certain number of Ramsey measurements using the same setup of $\tau_\text{{min}} = \pi/\omega_{\text{max}}$ and $\varphi_0\ = 0$. This is because the measurement outcome in Eq. (\ref{equ:1}) is a periodic signal, to avoid ambiguity, $\omega \tau + \varphi \leq \pi$ must hold. Therefore, for an unknown magnetic field in a wide range, we initiate the sensing time in such a conservative but non-optimal way. After collecting all the measurement results $\vec{\mu}$, the trained BNN estimator in the 1$^{st}$ Stage is employed to estimate the SoI $\omega$. The BNN estimator, initially proposed in~\cite{nolan2021frequentist}, is capable of learning a direct mapping from a sequence of measurement outcomes to the underlying parameter \(\omega\) without assuming any knowledge of the prior distribution. This data-driven approach offers several key advantages compared with the likelihood-based estimator. Specifically, it generalizes well across diverse scenarios, naturally handles rare outcomes, and supports fast online inference after being trained offline on simulated or experimental data. In light of it, when the optimal sensing parameters are yet unknown at the begining, BNN serves as a coarse-grained yet effective estimator to narrow down the $\omega$ for the second stage.

The output of the BNN estimator offers two fold of information --- the estimated value $\hat{\omega}$ and its associated uncertainty level $\varepsilon$, where \( \varepsilon=\mathbb{E}[|\hat{\omega} - \omega|]\). This output is subsequently used as the initial input of the protocol's 2$^{nd}$ Stage. Compared with the 1$^{st}$ Stage, the model in the 2$^{nd}$ Stage is adaptive and aims to optimize the sensing parameters \( (\tau, \varphi) \) over a small number of iterations. Specifically, based on the initial estimate \( (\hat{\omega}, \varepsilon) \) obtained from the 1$^{\text{st}}$ Stage, we construct an ensemble of candidate magnetic field values \( \{\omega_p\}_{p=1}^P \) uniformly distributed over the interval $[ \hat{\omega} - \frac{\Delta}{2}, \hat{\omega} + \frac{\Delta}{2} ]$, where \( 2\varepsilon \le \Delta < \omega_{\text{max}} \). This interval is selected to ensure, with high confidence, that the true field value lies within this range.

To approximate the posterior distribution \( p(\omega \mid \vec{\mu}^{(k)}) \), we employ a particle filter method, in which the posterior is represented as a weighted sum of Dirac $\delta$- functions centered at the particle locations:
\begin{equation}
    p(\omega \mid \vec{\mu}^{(k)}) = \sum_{p=1}^{P} a_p^{k} \cdot \delta(\omega - \omega_p).\label{equ:particle}
\end{equation}
The particle weights are initially set uniformly as \( a_p^0 = \frac{1}{P} \), reflecting a non-informative prior over the selected interval. Once the particle ensemble is initialized, the global RL agent begins its operation. At each sensing iteration \( k \), the agent selects a new control action \( (\tau_k, \varphi_k) \), which defines the sensing time and the phase of the control pulse, respectively. A single-shot measurement is then performed using this control setting, yielding a binary outcome \( \mu \in \{0, 1\} \). The particle weights are updated according to Bayes’ rule:
\begin{equation}
a_p^{k} = {a_p^{k-1} \cdot p(\mu \mid \omega_p)},  \label{equ:weight}
\end{equation}
where \( p(\mu \mid \omega_p) \) is the likelihood of observing \( \mu \) under the assumption that the true magnetic field is \( \omega_p \).

This closed-loop cycle—comprising RL-based control selection, single-shot readout, and posterior update—is repeated iteratively. The RL agent adapts its sensing strategy based on the current posterior to maximize information gain. The process continues until either the total sensing time budget is exhausted or the agent’s loss function converges below a predefined threshold. At the end of the 2$^{\text{nd}}$ Stage, a refined estimate of the magnetic field is computed as the weighted average of the particles:
\begin{equation}
\hat{\omega}_{\text{final}} = \sum_{p=1}^P a_p^K \cdot \omega_p,
 \label{equ:final_estimation}
\end{equation}
which improves upon the initial estimate \( \hat{\omega} \) and brings it closer to the true field value.
\subsection{Evaluation Results}
To evaluate the performance of the proposed method, we consider a single-spin system based on a NV center in diamond with single-shot readout capability~\cite{bonato2016optimized}. Key system parameters are summarized as follows. The spin coherence time is \( 96\,\mu\mathrm{s} \). Each Ramsey measurement includes a sensing time \( \tau \) and a system overhead of \( 240\,\mu\mathrm{s} \), which accounts for state preparation and readout. The total time budget for the magnetic sensing task is \( R_{\text{max}} = 22{,}000\,\mu\mathrm{s} \), consistent with the real-time constraints of applications such as geo-localization and medical diagnostics. The results presented in subsequent sections follow the windowing approach introduced in~\cite{belliardo2024model}.

In our analysis, we simulate 1,024 independent test trajectories over 200 iterations, yielding a total of \( 1,024 \times 200 \) samples. Each sample \( i \) corresponds to a true parameter value and proceeds through a sequence of \( M_i \) Ramsey measurements until the time budget is exhausted. For each sample \( i \), we record the used time resources and the corresponding estimation accuracy after the \( m \)-th Ramsey measurement as a tuple \( (t^{i}_m, \mathrm{MSE}^{i}_m) \), where \( t^{i}_m \) denotes the cumulative time consumed and \( \mathrm{MSE}^{i}_m \) the mean squared error at that step. If the final measurement exceeds the predefined time budget, its time and MSE are not recorded, to ensure the total resource usage remains within the specified constraint. We adopt a ``windowing approach'' which refers to a binning method that aggregates the tuples \( (t^{i}_m, \mathrm{MSE}^{i}_m) \) across all samples by discretizing the time axis. Specifically, we divide the time axis into bins of width \( 500\,\mu\mathrm{s} \), and assign each tuple to its respective bin according to $\left\lfloor {t^{i}_m}/{500\,\mu\mathrm{s}} \right\rfloor$. Within each bin, we compute the average time resources and corresponding MSE values. The resulting data points in the performance plots represent these mean values, offering a smoothed and interpretable visualization of the estimation accuracy as a function of consumed time resources. The line in the following figures simply connects the empirical data points to aid visual interpretation.



\begin{figure}[htbp]
	\centering
	\begin{subfigure}[b]{0.48\textwidth}
		\includegraphics[width=\linewidth]{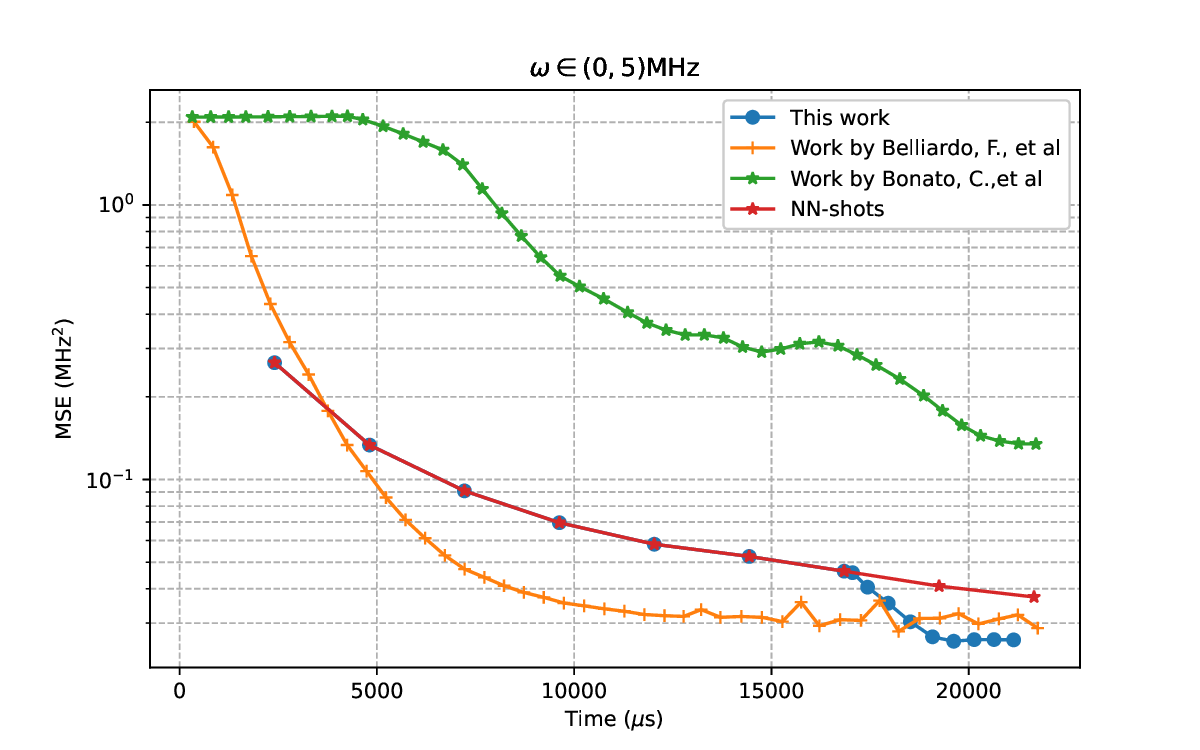}
		\caption*{(a)}
	\end{subfigure}
	\begin{subfigure}[b]{0.48\textwidth}
		\includegraphics[width=\linewidth]{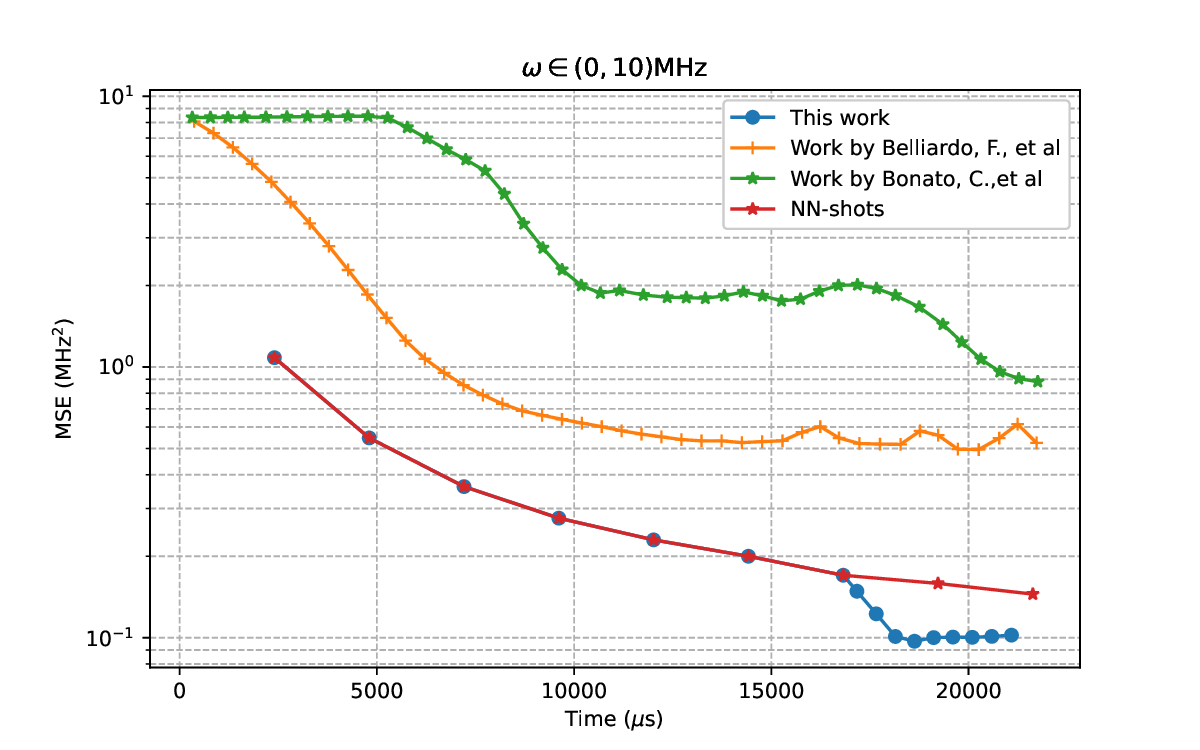}
		\caption*{(b)}
	\end{subfigure}
	\caption{Comparison of estimation performance across different methods. (a) shows the performance under the magnetic field range $\omega \in (0, 5)$ MHz. (b) illustrates the performance under the range $\omega \in (0, 10)$ MHz. In all cases, the non-adaptive 1$^{st}$ Stage consists of 70 measurement shots. The total available time resources are limited to $22,000~\mu$s, with a per-shot overhead of $240~\mu$s for spin initialization and readout, following Ref. \cite{bonato2016optimized}.  The curves for the protocols proposed by Belliardo et al. \cite{belliardo2024model} and Bonato et al. \cite{bonato2016optimized} begin after the first measurement shot, while the curves for NN-shots \cite{nolan2021machine} and the protocol proposed in this work begin after the initial 10 measurement shots.}  
	\label{fig:omega5:10}
\end{figure}

\begin{figure}[h]
	\centering
	\includegraphics[width=0.9\textwidth]{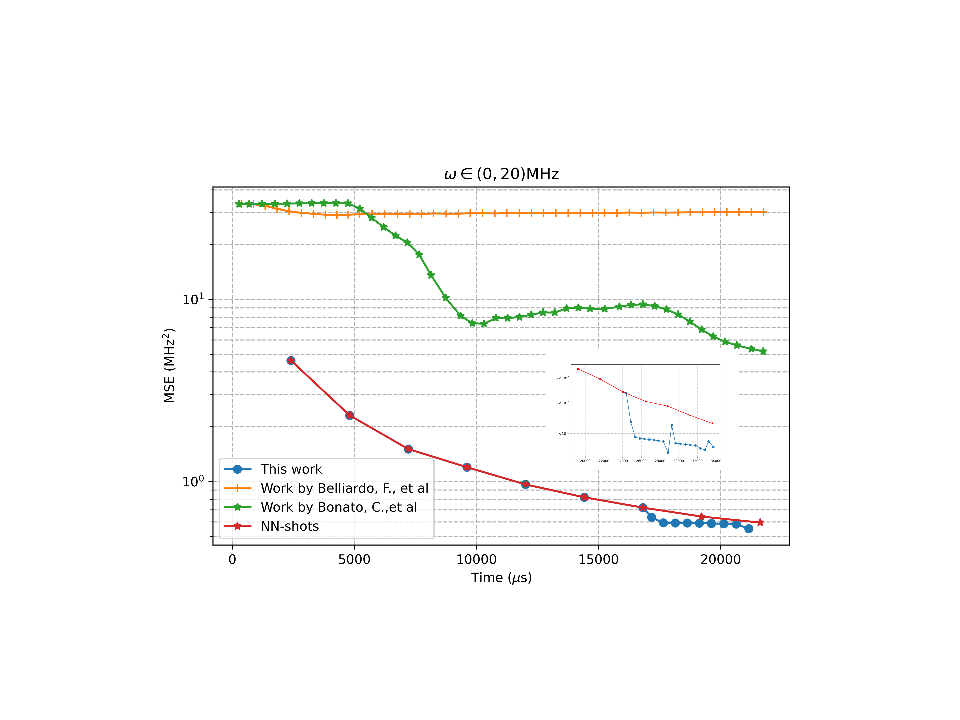}
	\caption{Comparison of estimation performance under a wide magnetic field range $\omega \in (0, 20)$ MHz. The non-adaptive 1$^{st}$ Stage consists of 70 measurement shots. The inset figure is a zoom-in view of the performance of the proposed protocol compared with that of the NN-shots method in \cite{nolan2021machine} when the number of measurement shots in the non-adaptive 1$^{st}$ Stage is increased to 100.}\label{fig:omega20}
\end{figure}

Fig. \ref{fig:omega5:10} presents the performance of various sensing protocols under two magnetic field ranges: $\omega \in (0, 5)$ MHz and $\omega \in (0, 10)$ MHz. 
The results demonstrate that the proposed protocol in this work outperforms existing methods in terms of estimation accuracy under limited time resources. In Fig. \ref{fig:omega5:10}(a), under the magnetic field range of $\omega \in (0, 5)$ MHz, the proposed method outperforms the NN-shots approach of Nolan et al.~\cite{nolan2021machine} by $30.2\%$, and the method of Belliardo et al.~\cite{belliardo2024model} by $9.5\%$. The method by Belliardo et al. also shows a $22.9\%$ improvement over the NN-shots approach. In Fig. \ref{fig:omega5:10}(b), within the broader field range of $\omega \in (0, 10)$ MHz, the proposed method achieves a $29.5\%$ performance improvement over the NN-shots method. In Fig. \ref{fig:omega5:10}(a) where $\omega$ is in a narrower range, the RL-based method by Belliardo et al. \cite{belliardo2024model} shows better performance than the methods by Bonato et al. \cite{bonato2016optimized} and Nolan et al. \cite{nolan2021machine}. This is attributed to the adaptive nature of RL that optimizes $\{\tau, \varphi\}$ collectively; while the other two methods are either non-adaptive \cite{nolan2021machine} or have smaller optimization space \cite{bonato2016optimized}. Interestingly, when the range of $\omega$ broadens, Fig. \ref{fig:omega5:10}(b) shows that although the RL-based method by Belliardo et al. \cite{belliardo2024model} is still better than the method by Bonato et al. \cite{bonato2016optimized}, it underperforms the NN-shots method by Nolan et al. \cite{nolan2021machine}. This suggests that the RL agent trained in \cite{belliardo2024model} lacks the capability required for effective optimization across a much wider solution space (i.e., a wider field range). In brief summary, the lessons learned from these existing methods are two-fold: (1) the vanilla BNN estimator using non-adaptive parameter setup is ideal for estimating $\omega$ in a wide range; (2) the RL method is well-suited to optimize sensing parameters in a small search space but fails to generalize when the space explodes. To combine the benefits of the two methods, this work's approach includes running the BNN estimator with fixed parameter setup for one-shot estimation and further utilizing a federated RL agent to further optimize the parameters within a subrange of $\omega$. As shown in Fig. \ref{fig:omega5:10}(a-b), our method manages to boost the performance beyond the best among the state of the art. 

Fig. \ref{fig:omega20} further shows the performance of same baseline protocols and ours in an extended magnetic field range of $\omega \in (0, 20)$ MHz. 
The results indicate that our method achieves superior estimation accuracy compared to others, given the same limited time resources ($R_{\text{max}} = 22{,}000~\mu$s). The proposed method demonstrates a $7.5\%$ performance improvement over the NN-shots method of Nolan et al.~\cite{nolan2021machine}. Furthermore, when the number of measurement shots in the 1$^{st}$ Stage is increased to 100, the inset figure demonstrates that the advantage becomes more pronounced, with the proposed method outperforming the NN-shots approach \cite{nolan2021machine} by $17.6\%$. These results confirm that the RL agent in the 2$^{nd}$ Stage of our protocol is capable of optimizing sensing parameters to enhance estimation accuracy and efficiency.
In addition, it is observed that the estimation accuracy of the protocol by Bonato et al. \cite{bonato2016optimized} exceeds that of Belliardo et al. \cite{belliardo2024model} in this large field range. Recall that the method in \cite{bonato2016optimized} employs a partially adaptive strategy --- only phase $\varphi$ is adaptively optimized while $\tau$ is selected from a predetermined set. The method in \cite{belliardo2024model} optimizes both $\varphi$ and sensing time $\tau$. Fig. \ref{fig:omega20} reveals that the latter method does not give any edge. This comparison highlights the potential advantage of fixing sensing time $\tau$ (ideally equal to $\tau_{\text{min}}$) in the early stage of the protocol to save the RL agent from iterating over unnecessarily long steps for the search of an optimal parameter setup. This design principle is particularly helpful when the total sensing time as a resource is limited and meanwhile the range of the magnetic field is large.

Moreover, we understand that the convergence rate and optimality in the vanilla RL framework is cursed by large search space. To enhance its performance for wide-range magnetic field sensing, we introduce federated learning into the training process of the RL agent used in the 2$^{nd}$ Stage. 
As shown in the offline training phase for the 2$^{nd}$ Stage in Fig. \ref{fig:summary}, the entire magnetic field range is partitioned into multiple intervals. Each interval is treated as a local environment, corresponding to a specific subrange of the D.C. magnetic field. A dedicated local RL agent is trained within each of these environments, producing local updates to the neural network parameters. These local updates are then aggregated using the FedAvg algorithm to update a global RL agent. Importantly, the local environments remain fixed throughout the training process, ensuring that each local agent learns to optimize sensing parameters for the $\omega$ within its designated subrange.
The posterior probability distribution, i.e., the Bayesian update of the estimated D.C. magnetic field, is used as the input to the global RL agent. This distribution is approximated using a particle filter implemented via a sequential Monte Carlo (SMC) algorithm. During the online inference phase, the range of the particles is set to the subrange of $\omega$ identified based on the estimation value $\hat{\omega}$ and associated uncertainty $\varepsilon$ obtained from the 1$^{st}$ Stage. This innovation, which we call federated RL agent, plays a vital role for the outperforming results of our protocol shown in the earlier figures. 
To support the robustness and statistical reliability of our numerical results, we conducted a detailed analysis of the estimation error and corresponding confidence intervals under various conditions in Supplementary Section 4 of the Supplemental document.
\begin{figure}[htbp]
	\centering
	\begin{subfigure}[b]{0.32\textwidth}
		\includegraphics[width=\linewidth]{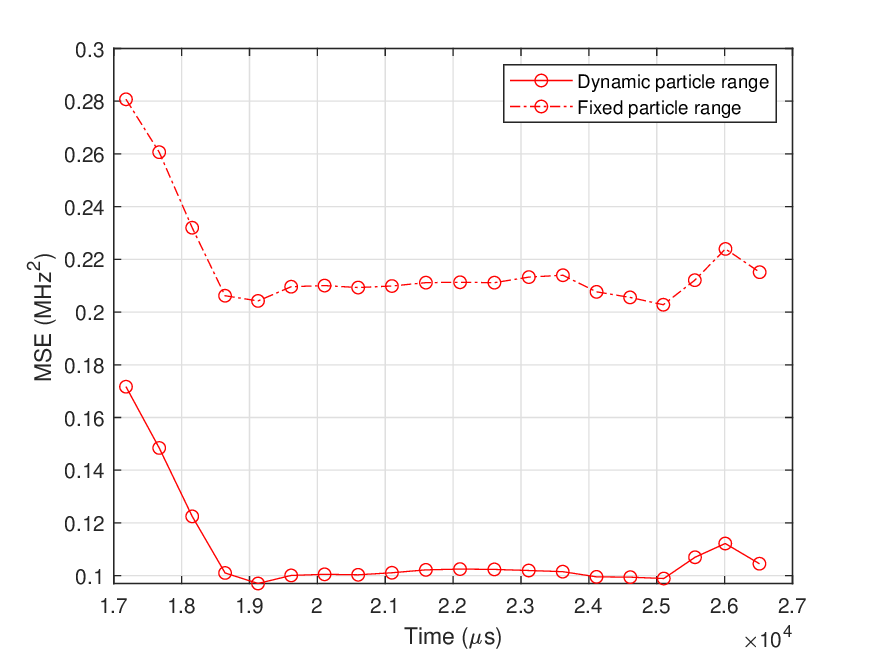}
		\caption*{Federated-RL}
	\end{subfigure}
	\begin{subfigure}[b]{0.32\textwidth}
		\includegraphics[width=\linewidth]{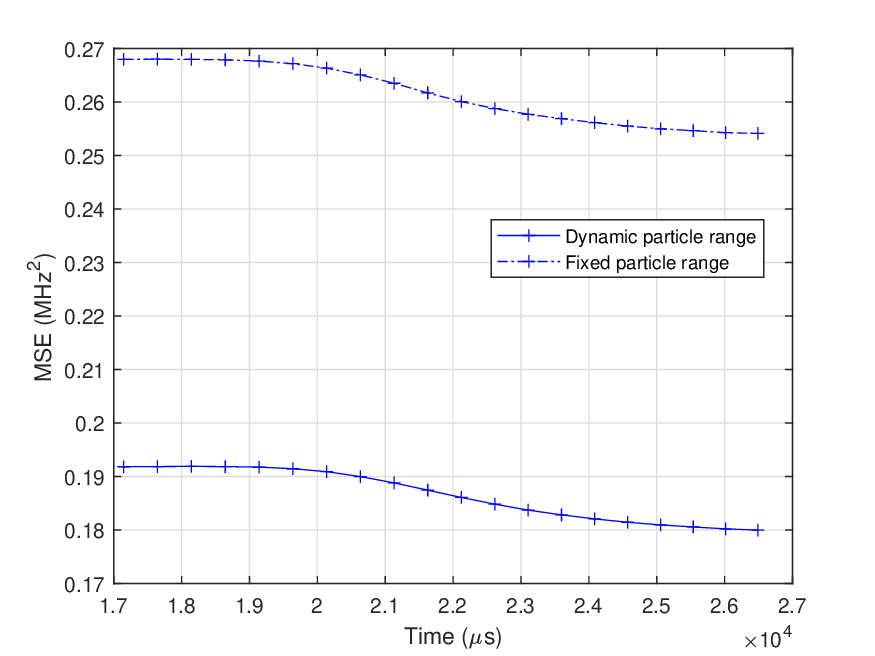}
		\caption*{oneM-RL}
	\end{subfigure}
	\begin{subfigure}[b]{0.32\textwidth}
		\includegraphics[width=\linewidth]{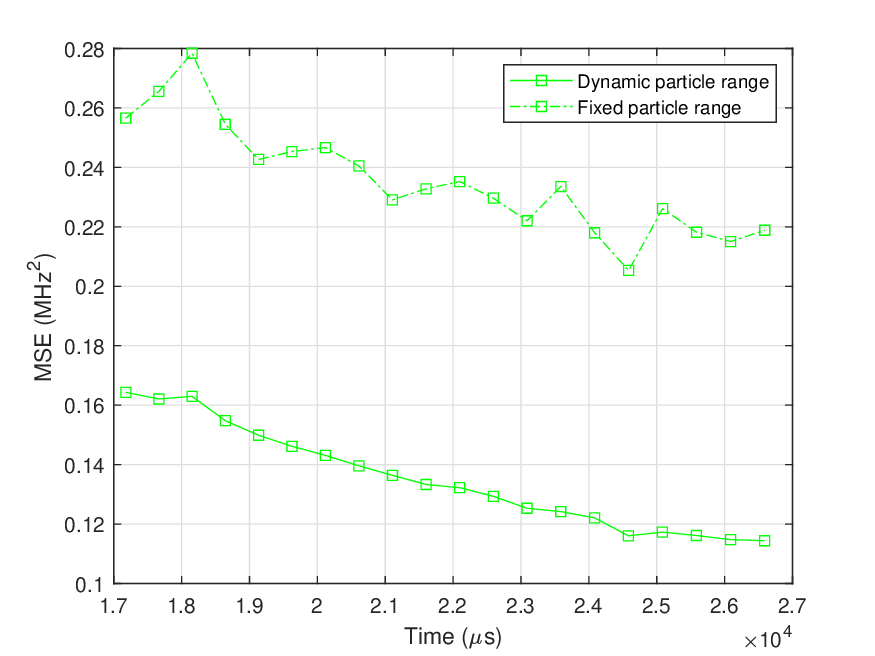}
		\caption*{multipleM-RL}
	\end{subfigure}
	\caption{Comparison of different strategies for setting the particle range. \texttt{oneM-RL} stands for one single RL \underline{M}odel while \texttt{multipleM-RL} represents many parallel RL \underline{M}odels. The results are collected for $\omega \in (0,10)$MHz.}
	\label{fig:delta:selection}
\end{figure}

\begin{figure}[htbp]
	\centering
	\begin{subfigure}[b]{0.31\textwidth}
		\includegraphics[width=\linewidth]{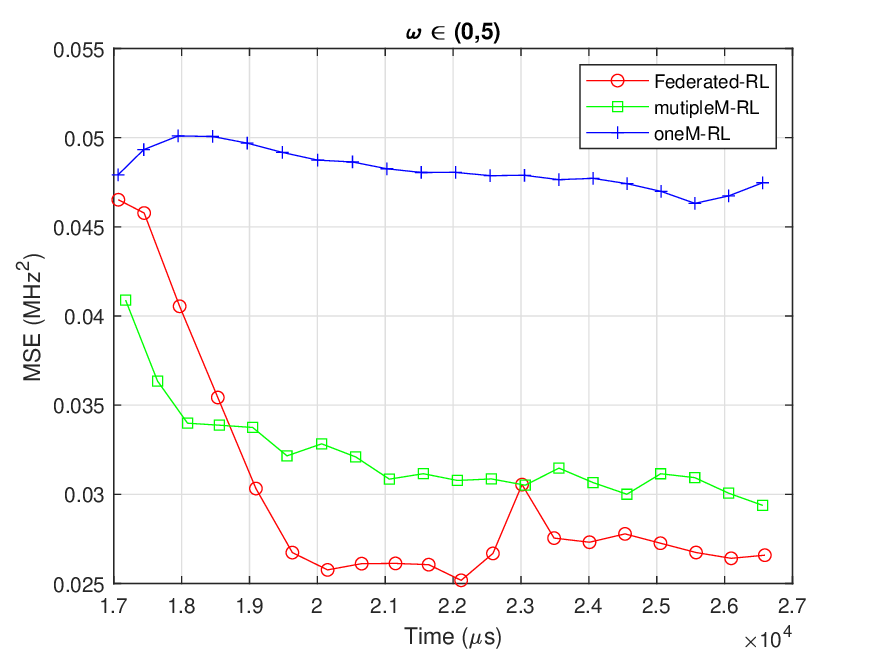}
		\caption*{(a)}
	\end{subfigure}
	\begin{subfigure}[b]{0.31\textwidth}
		\includegraphics[width=\linewidth]{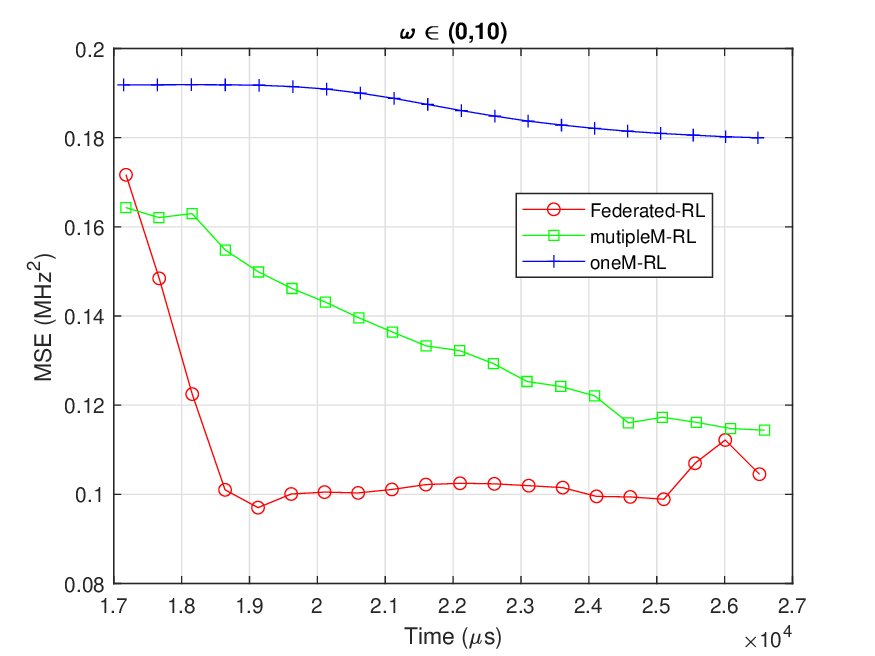}
		\caption*{(b)}
	\end{subfigure}
	\begin{subfigure}[b]{0.31\textwidth}
		\includegraphics[width=\linewidth]{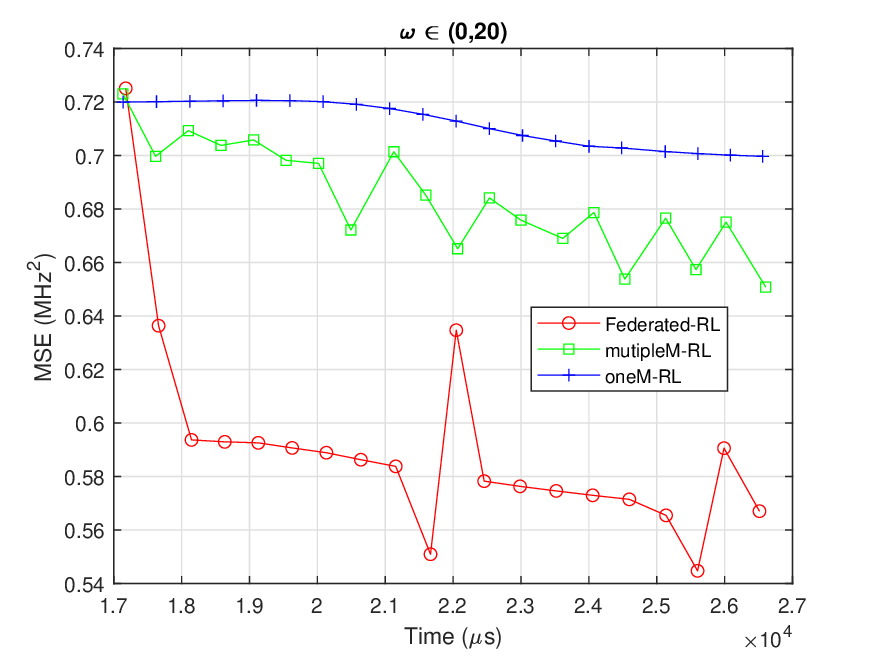}
		\caption*{(c)}
	\end{subfigure}
	\vspace{5pt} 
	\begin{subfigure}[b]{0.31\textwidth}
	\includegraphics[width=\linewidth]{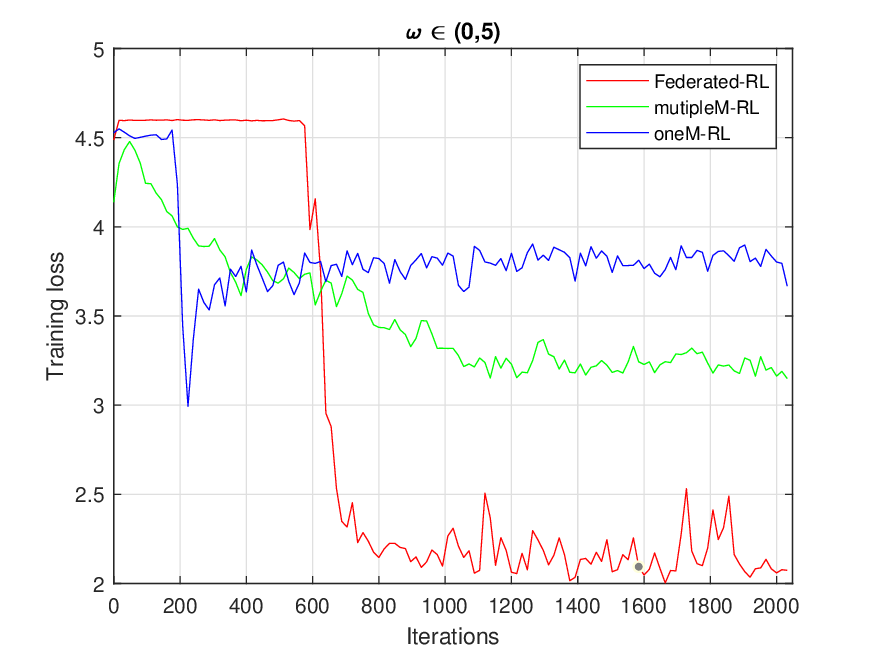}
	\caption*{(d)}
\end{subfigure}
\begin{subfigure}[b]{0.31\textwidth}
	\includegraphics[width=\linewidth]{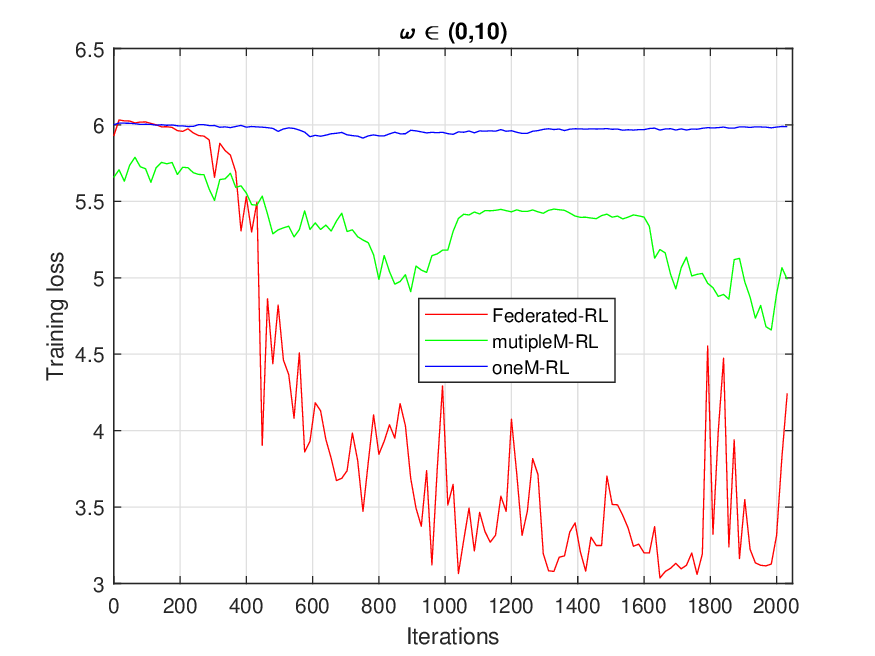}
	\caption*{(e)}
\end{subfigure}
\begin{subfigure}[b]{0.31\textwidth}
	\includegraphics[width=\linewidth]{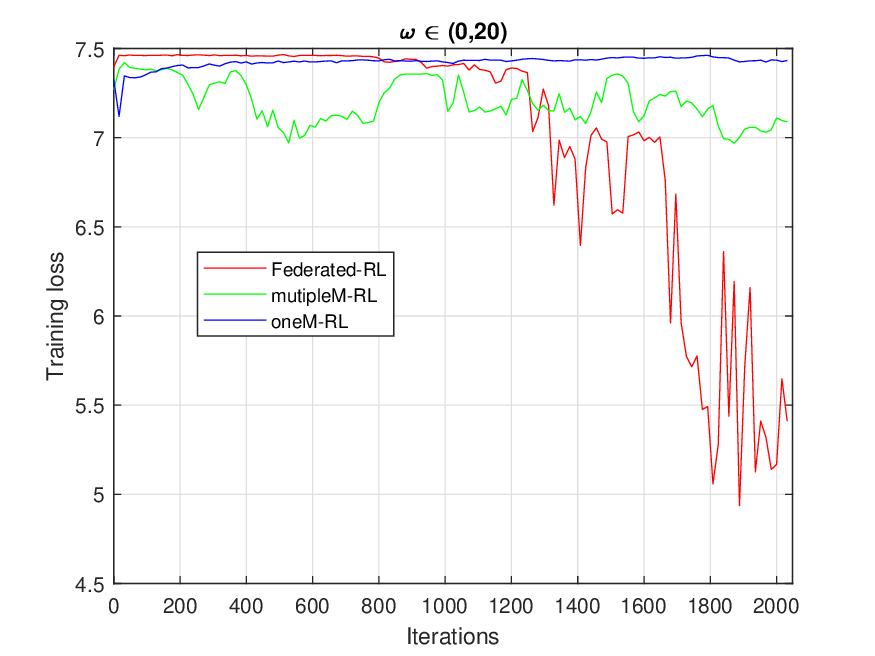}
	\caption*{(f)}
\end{subfigure}
	\caption{Comparison of performance among the three RL agents with the dynamic particle range. (a)–(c) show the estimation accuracy of the 2$^{nd}$ Stage following 70 measurement shots in the 1$^{st}$ Stage; and (d)-(f) compare their convergence rate and training loss.}
	\label{fig:separate:stage}
\end{figure}

To further highlight the advantages of the federated RL agent, we compare its standalone performance with two additional RL training strategies in the following experiments. The results are shown in Fig. \ref{fig:delta:selection} and Fig. \ref{fig:separate:stage}. Here, \texttt{Federated-RL} refers to the proposed method used in this work, in which a global RL agent is trained via the FedAvg algorithm by aggregating updates from multiple local agents, each trained within a specific subrange of the magnetic field. \texttt{oneM-RL} denotes a sequential training approach, where a single RL agent is trained across all local environments one after another. \texttt{multipleM-RL} represents a scheme in which multiple RL agents are independently trained within their respective local environments, without any aggregation or parameter sharing. This comparison aims to demonstrate that our proposed method not only achieves superior generalization across the wide magnetic field range but also maintains the efficiency and adaptability needed for high-accuracy estimation. Note that the notion of generalization considered in this work differs slightly from that typically discussed in the context of federated learning. Here, generalization refers to the ability of the global RL agent to outperform individual local agents across the entire estimation range. While each local RL agent is trained to identify an optimal experimental design within its assigned subrange, the global RL agent integrates information from all subranges and is thus capable of learning a more general policy that performs well across the full estimation domain. 
The evaluation of the three methods is conducted under two distinct setups: one employing a \textit{dynamic particle range} defined around the estimate obtained from the first stage, and the other using a \textit{fixed particle range} corresponding to predefined intervals from the training phase.
The key difference between these two approaches lies in how the particle interval is constructed based on the BNN estimate from the first stage. For the dynamic particle range approach, given the first-stage estimate \(\hat{\omega}\), the particle range is defined as a symmetric interval centered on \(\hat{\omega}\), i.e., \(\left[\hat{\omega} - \Delta/2,\, \hat{\omega} + \Delta/2\right]\) with $\Delta$ selected according to the principle in Section~\ref{sec12}. This adaptive localization enables the second stage to focus computational resources on regions where the true parameter is most likely to reside, thereby enhancing estimation efficiency. In the fixed particle range approach, the global frequency domain \([0, \omega_{\max}]\) is uniformly partitioned into equal-width subintervals. Once a frequency estimate \(\hat{\omega}\) is obtained, the second-stage particle filter is initialized within the specific subinterval that contains \(\hat{\omega}\). 
 While simpler to implement, this approach risks placing particles in regions that may not fully capture the posterior distribution, especially if \(\hat{\omega}\) lies near a subinterval boundary. In this work, the number of fixed subintervals is set as 10. As shown in Fig.~\ref{fig:delta:selection}, the dynamic particle range setup consistently outperforms the fixed particle range approach across all three models. This improvement arises because the true value of the D.C. magnetic field is more likely to be located near the first-stage estimate, enabling the dynamic range to more effectively concentrate particles where they are most needed.

 More importantly, this work's \texttt{Federated-RL} framework outperforms the other two RL frameworks in terms of MSE, almost for any given sensing time budget in the x-axis. This performance advantage is theoretically supported by the well-known strengths of federated learning in non-independent and identically distributed (non-IID) settings, where centralized training often suffers from destructive gradient interference. By avoiding direct data mixing, the \texttt{Federated-RL} approach stabilizes training and enables robust policy learning across structurally diverse subdomains. The results presented in Fig.~\ref{fig:delta:selection} provide strong empirical justification for adopting the \texttt{Federated-RL} framework in our work.

Next, Fig. \ref{fig:separate:stage} compares the performance of the three RL agents under different magnetic field ranges: $\omega \in (0, 5)$ MHz, $\omega \in (0, 10)$ MHz, and $\omega \in (0, 20)$ MHz. The results demonstrate that the proposed federated RL agent consistently achieves significantly higher estimation accuracy and efficiency across all tested field ranges. This performance advantage is consistent with the training loss curves shown in Fig. \ref{fig:separate:stage}(d)–(f), confirming that our approach achieves much lower training loss while maintaining a similar level of convergence speed. In Supplementary Section 5 of the Supplemental document, we further demonstrate the real-time evolution of sensing parameters as the adaptive 2$^{nd}$ Stage proceeds. This result will offer a granular view of the trajectory of sensing parameter evolvement, elucidating the reason why our protocol achieves superior performance over the methods by Belliardo et al.~\cite{belliardo2024model} and Bonato et al.~\cite{bonato2016optimized}.

\section{Methods}\label{sec11}
\subsection{Protocol Details}
The framework of the proposed two-stage optimization method is illustrated in Fig. \ref{fig:summary}. In the non-adaptive 1$^{st}$ Stage, a BNN estimator is employed to narrow the range of the estimated magnetic field. The BNN takes the measurement outcomes from a sequence of Ramsey experiments as input and outputs an estimated magnetic field value. To avoid ambiguity arising from the periodic nature of the Ramsey signal, the sensing time is fixed at $\tau_{\text{min}} = \pi/\omega_{\text{max}}$ and the control phase $\varphi_0 = 0$. Details of the BNN training process are provided in Supplementary Section 1 of the Supplemental document.

In the adaptive 2$^{nd}$ Stage, an RL agent is employed to optimize the sensing time and control phase for estimating the magnetic field within a subrange centered around the value obtained from the 1$^{st}$ Stage. This subrange is defined as an interval with a width slightly greater than twice the estimated error. However, training a single RL agent over a broad and complex domain presents a challenging optimization task, particularly when using function approximates such as deep neural networks. This often results in slower learning, reduced stability, and increased susceptibility to local minima. By decomposing the global task into a set of simpler, localized sub-problems during the offline training phase, federated learning allows each local agent to optimize within a narrower and smoother loss landscape, facilitate faster convergence and improved solution quality while obtaining a global agent to generalize effectively across all possible subranges. The details of the federated RL agent training process are provided in Supplementary Section 2 of the Supplemental document. Specifically, a key input to the federated RL agent is the posterior distribution of the estimated magnetic field, which is approximated using a particle filter.
The posterior distribution \( p(\omega \mid \vec{\mu}^{(k)}) \) is represented using a particle filter, where \( \vec{\mu}^{(k)} \) denotes the sequence of measurement outcomes up to the \( k \)-th Ramsey experiment. According to Bayes' theorem, the posterior is recursively updated via
\begin{equation}
p(\omega \mid \vec{\mu}^{(k)}) \propto p(\mu \mid \omega) \cdot p(\omega \mid \vec{\mu}^{(k-1)}), \label{equ:posterior}
\end{equation}
where \( \mu \) is the latest measurement outcome. 

During the adaptive 2$^{nd}$ Stage, each iteration consists of three steps. First, the last round of Ramsey measurement produces a measurement outcome $\mu$ that results in the update of posterior probability in Eq. (\ref{equ:posterior}) and its associated particle weight in Eq. (\ref{equ:weight}). Then, the posterior probability is approximated by the particle filter method following Eq. (\ref{equ:particle}). It is further combined with the remaining time resources as the input to the federated RL agent, which then outputs an optimized sensing time and controlled phase for the next Ramsey measurement. Second, a Ramsey experiment is performed using these optimized sensing parameters to obtain a new measurement outcome. Third, the posterior distribution is updated again according to Bayes’ rule in Eq. (\ref{equ:posterior}), and each particle is reweighted based on the likelihood of the new measurement result as in Eq. (\ref{equ:weight}). This process is repeated until the total available time resources are exhausted.
Details regarding the computational overhead associated with both the BNN and the federated RL components are provided in Supplementary Section 3 of the Supplemental document.

\section{Discussion}\label{sec12}
Several key directions about the proposed two-stage optimization method warrant further investigation to enhance its practical performance and applicability.

Analysis of \( \Delta \): The choice of \( \Delta \) should take into account the estimation error introduced during the adaptive stage. This error arises from a combination of factors, including the training error, the size of the measurement vector $\vec{\mu}$, the dimensionality of the BNN’s output layer (reflecting the granularity of the discretization of $\omega$), the posterior approximation error, and others. From existing knowledge \cite{lehmann2006theory}, we understand that in an ideal situation where the BNN estimator is sufficiently well-trained, the output Bayesian posterior probability shall asymptotically converges to the Gaussian distribution as the number of measurements $M$ in $\vec{\mu}$ increases:
\begin{equation}
p(\omega_j | \vec{\mu}) \approx \sqrt{\frac{M F(\omega)}{2\pi}} \, e^{- \frac{M F(\omega)}{2} (\omega_j - \omega)^2} \label{eq:gaussian}
\end{equation}
The distribution is centered at the true value $\omega$ and with variance $1/[M F(\omega)]$ where $F(\omega)$ is the Fisher information. However, selecting the confidence level $\Delta$ based on the standard deviation from Eq. (\ref{eq:gaussian}) may not be very trustworthy in a practical setting. Therefore, to ensure that the adaptive stage operates within a reliably estimated subrange, the chosen discretization interval \( \Delta \) must satisfy the constraint \( 2\varepsilon \le \Delta < \omega_{\text{max}} \). In this work, we observe that the corresponding BNN estimation errors after 70 Ramsey measurements are approximately \( 0.22 \), \( 0.41 \), and \( 0.85\,\text{MHz} \) for magnetic field ranges \( \omega \in (0, 5) \), \( (0, 10) \), and \( (0, 20)\,\text{MHz} \), respectively. Accordingly, \( \Delta \) is empirically chosen as \( 0.5 \), \( 1 \), and \( 2\,\text{MHz} \) for these ranges.
The principle guiding the choice of \(\Delta\) is illustrated and justified by simulation results provided in Supplementary Section 6 of the Supplemental document.

While the chosen \( \Delta \) performs well within the protocol setting of this work, we acknowledge that it can be further reduced or dynamically adjusted at runtime without increasing the complexity of the BNN model or the number of measurement samples (which otherwise results in waste of time resources). Such optimization has the potential to further enhance the overall estimation accuracy.


Time Resource Allocation: A practical quantum sensing system may operate under limited time budget due to the latency requirement of a sensing task or the system overhead. In this work, we follow the experimental setup in Bonato et al. \cite{bonato2016optimized} and set the time budget be $R_{\text{max}} = 22{,}000~\mu$s. We spare 70 measurement shots to the 1$^{st}$ Stage while allocating the rest to the 2$^{nd}$ Stage. In this configuration, we in fact allocate \( \left(t_\text{inactive} + {\pi}/{\omega_{\text{max}}}\right) \times 70 \mu\)s time resources to the 1$^{st}$ Stage when accounting for the inactive time $t_\text{inactive}$ spent for preparing, reading out, and resetting qubit for each shot of measurement. Following the setup in~\cite{bonato2016optimized}, we let $t_\text{inactive}$  be 240 $\mu$s. In an ideal situation that time resource is not a constraint, by accommodating 100 measurement shots in the 1$^{st}$ Stage while keeping the same time for the 2$^{nd}$ Stage, the inset figure in Fig. \ref{fig:omega20} reveals that the accuracy can be further boosted by 10\%. However, in a realistic time-limited context, the resource allocation between the two inference stages must be carefully planned. We believe that further analysis of \( \Delta \) mentioned in the prior section and exploration of the parallelism of the two stages will shed the light on the pathway for time resource optimization, leading to further improvement on estimation accuracy. 

Scalability Beyond NV Centers: While our protocol is demonstrated using a single NV center in a diamond as the quantum sensor platform, the optimization framework is inherently sensor-agnostic. The BNN estimator only requires training data comprising measurement outcomes and does not depend on sensor-specific physics. Similarly, the RL agent is trained to optimize sensing parameters (e.g., time, phase) and can be adapted to other sensor platforms by redefining the action space. For example, the protocol can be extended to systems beyond the single-shot readout capability by treating the Bayesian update input as a vector \( \vec{\mu} \) of aggregated outcomes. This flexibility enhances its applicability to a broad class of quantum sensing and metrology systems.

Robustness to Noise: A significant advantage of the proposed method is the separation of offline training and online inference. Both the BNN estimator and the RL agent can be trained on large, noisy datasets, allowing them to learn noise-robust representations and policies. Furthermore, federated learning enables the RL agent to aggregate updates from local environments with varied noise characteristics, resulting in a globally trained agent with improved generalization to practical sensing scenarios. For instance, if considering the measurement noises, we can update Eq. (\ref{equ:1}) with $p(0 \mid \omega) = \frac{1 + F_0 - F_1}{2} + \frac{F_0 + F_1 - 1}{2} e^{-\tau / T_2} \cos(\omega \tau + \varphi)$ where $F_0$ and $F_1$ are readout fidelity for 0s and 1s respectively. Further simulation results under noisy conditions are provided in Supplementary Section 4 of the Supplemental document to demonstrate the robustness of the proposed scheme to measurement noise.
By profiling the noise models of a quantum sensing system at design time, the inference framework can be adjusted offline without introducing any runtime overhead. This flexibility offers great robustness to noisy intermediate-scale quantum (NISQ) devices.

Therefore, the proposed two-stage optimization method offers a scalable, efficient, and noise-robust solution for various quantum sensing apparatus. To draw a conclusion, this work presented a new quantum magnetic sensing protocol for a NV center system with only single-shot readout capability. The protocol introduces a novel progressive optimization strategy that led to a superior performance over the state-of-the-art methods, most notably a 7.5\% accuracy improvement for estimating a wide-range magnetic field within a constrained sensing time budget of less than 22 ms. Furthermore, our analysis showed that the protocol's framework is both scalable to other quantum sensing platforms and robust against various noise sources. Future work will explore optimal time allocation strategies, adaptive tuning of \( \Delta \), and applications to other sensor platforms, further enhancing its utility for precision metrology in diverse experimental settings.
\backmatter

\bmhead{Supplementary information}
The online version contains supplementary material available at [URL to be filled].

\bmhead{Acknowledgments}
J. Liu discloses support for the research of this work from National Science Foundation [grant number 2304118 and 2326746].

\bmhead{Competing Interests}
The authors declare no competing financial or non-financial interests.

\section*{Declarations}
\begin{itemize}
\item Data availability: The datasets generated during and/or analysed during the current study are available
from the corresponding author on reasonable request.

\item Code availability: The underlying code for this study is not publicly available but may be made available to qualified researchers on reasonable request from the corresponding author.

\end{itemize}

\clearpage
\newpage
\begin{center}
    \textbf{Supplementary Material: A Two-stage Optimization Method for Wide-range Single-electron Quantum Magnetic Sensing}
\end{center}
\setcounter{section}{0}
\section{Training of the BNN Estimator}\label{secMethods-bnn}
The BNN estimator is implemented as a fully connected feed-forward neural network comprising two hidden layers, each with 32 neurons. The hidden layers employ the ReLU activation function, defined as $\max(x, 0)$. The input and output dimensions of the network are 1 and 100, respectively. We encode one training sample (or measurement outcome) from the dataset $\mathcal{D}_\omega$ as a one-hot vector and feed into the input layer of BNN. The output layer is normalized using the softmax function to produce a valid probability distribution over discrete magnetic field values. Specifically, the network outputs a posterior distribution $p(\omega_j \mid \mu_i)$, which represents the probability of the magnetic field taking the value $\omega_j$ conditioned on a single-shot measurement outcome $\mu_i$. 

The loss function is defined as the average negative log-likelihood across all training samples:
\begin{equation}
\mathcal{L} = -\frac{1}{b_\omega |\mathcal{D}_\omega|} \sum_{\ell = 1}^{b_\omega} \sum_{i = 1}^{|\mathcal{D}_\omega|} \log p(\omega_\ell \mid \mu_{i,\ell}),
\end{equation}
where $b_\omega$ denotes the number of discrete values within the range $\omega_\ell \in (0, \omega_{\text{max}})$, and $|\mathcal{D}_\omega|$ is the number of measurement outcomes in the dataset. To improve generalization and training stability, an $L_2$ regularization term is added to the loss. Optimization is performed using the Adam algorithm~\cite{diederik2014adam} for stochastic gradient descent. All neural network estimators are implemented and trained using the JAX framework~\cite{bradbury2018jax}.

Assuming conditional independence of measurements and a uniform prior over $\omega$, the posterior distribution over $\omega_j$ given a sequence of $M$ measurement outcomes $\vec{\mu}^{(M)} = \{\mu_1, \mu_2, \dots, \mu_M\}$ is given by the normalized product of individual posteriors:
\begin{equation}
p(\omega_j \mid \vec{\mu}) \propto \prod_{m=1}^{M} p(\omega_j \mid \mu_m).\label{A2}
\end{equation}

Equation~(\ref{A2}) can be obtained via Bayesian updating. Specifically, the posterior distribution after incorporating the $M$-th measurement outcome $\mu_m$ is given by
\begin{equation}
	p(\omega_j \mid \vec{\mu}^{(M)}) = p(\mu_m \mid \omega_j) \cdot p(\omega_j \mid \vec{\mu}^{(M-1)}),
\end{equation}
where, according to Bayes’ theorem, the conditional probability $p(\mu_m \mid \omega_j)$ can be expressed as
\begin{equation}
	p(\mu_m \mid \omega_j) = \frac{p(\omega_j \mid \mu_m) \cdot p(\mu_m)}{p(\omega_j)}.
\end{equation}
Substituting into the update rule yields
\begin{equation}
	p(\omega_j \mid \vec{\mu}^{(M)}) = \frac{p(\omega_j \mid \mu_m) \cdot p(\mu_m)}{p(\omega_j)} \cdot p(\omega_j \mid \vec{\mu}^{(M-1)}).
\end{equation}
By iterating this update over successive measurements, one arrives at Eq.~(\ref{A2}). Here, $p(\omega_j)$ represents the prior distribution, and $p(\mu_m)$ serves as a normalization factor to ensure the posterior remains properly normalized. Thus, the posterior distribution shown in Figure 1 of the main document is given by
\begin{equation}
	p(\omega_j \mid \vec{\mu}^{(M)}) \propto p(\omega_j) \cdot \prod_{m=1}^{M} p(\omega_j \mid \mu_m).
\end{equation}
And the final estimation value is then obtained via maximum a posteriori (MAP) inference:
\begin{equation}
\hat{\omega} = \arg\max_{\omega_j} \left[ p(\omega_j \mid \vec{\mu}) \right].
\end{equation}

A training dataset for the BNN estimator consists of measurement outcomes, i.e., classical bit-strings, denoted as $\mu_{m} \in \{0, 1\}$. These outcomes are collected at $b_{\omega}$ distinct true parameter values uniformly sampled across the range $\left(0, \omega_{\text{max}}\right)$. For every Ramsey measurement, we always choose $\tau = \pi/\omega_{\text{max}}$ and $\varphi_0 = 0$ for quantum magnetic sensing. For each value of $\omega$, a sequence of $|\mathcal{D}_{\omega}|$ Ramsey measurements are performed and their  outcomes are recorded. Thus, the total number of bit-strings in the training dataset is $b_{\omega} \times |\mathcal{D}_{\omega}|$, forming a two-dimensional binary array of shape $\left(b_{\omega}, |\mathcal{D}_{\omega}|\right)$. The batch size is $b_\omega = 100$ and the dataset comprises $|\mathcal{D}_\omega| = 5,000$ samples. Training is performed over 8,000 iterations with a learning rate of $10^{-3}$.

\section{Training of the Federated RL Agent}\label{secMethods-frl}
The RL agent is implemented as a fully connected neural network comprising 5 hidden layers, each with 64 neurons. The input and output dimensions of the network are 5 and 2, respectively. A hyperbolic tangent (tanh) activation function is applied to the final layer to bound the output within the interval $[-1, 1]$.

The input (observation) to the agent after the $k^{\text{th}}$ Ramsey experiment is denoted as 
\begin{equation}
s_k = \left(\widetilde{\omega}_k, \widetilde{\sigma}_k, \Gamma_k, \widetilde{k}, \widetilde{R}_k\right),
\end{equation}
where all components are scaled to lie within the range $[-1, 1]$. Here, $\widetilde{\omega}_k$ represents the normalized mean of the estimation posterior, and
\begin{equation}
\widetilde{\sigma}_k = -\frac{2}{10} \ln \sqrt{\Sigma_k} - 1,
\end{equation}
with $\Sigma_k$ denoting the covariance (or variance in the scalar case) around the posterior mean, defined as
\begin{equation}
\Sigma_k := \int (\omega - \hat{\omega}_k)(\omega - \hat{\omega}_k)^\top p(\omega \mid \vec{\mu}^{(k)}) \, d\omega,
\end{equation}
The term $\Gamma_k$ represents the correlation matrix of the posterior distribution. For the single-parameter estimation problem considered here, this simplifies to $\Gamma_k = 1$. The variables $\widetilde{k}$ and $\widetilde{R}_k$ denote the normalized experiment index and normalized consumed time resources, respectively.

The agent, parameterized by $\boldsymbol{\lambda}$, outputs two actions: sensing time $\tau$ and controlled phase $\varphi$. This new sensing setup is then applied to the subsequent Ramsey measurement to refine the parameter estimation.

The local loss function, capturing the cumulative estimation error over a sequence of experiments, serves as the reward signal for training~\cite{fiderer2021neural}. For agent $n$ trained on dataset $\mathcal{D}_n$, the loss is defined as
\begin{equation}
\mathcal{L}_n(\boldsymbol{\lambda}) := \frac{1}{K |\mathcal{D}_n|} \sum_{k=1}^{K} \sum_{i=1}^{|\mathcal{D}_n|} \ell(\widehat{\omega}_{i,k}, \omega_i),
\end{equation}
where $K$ is the maximum number of Ramsey experiments under the limited time resource $R_{\text{max}}$, $\omega_i \in [n{\omega_\text{max}}/{N}, (n+1){\omega_\text{max}}/{N})$ denotes the true value and $\widehat{\omega}_{i,k}$ is the estimate after the $k^{\text{th}}$ Ramsey measurement. The estimate is computed as a weighted sum over basis points:
\begin{equation}
\widehat{\omega}_{i,k} = \sum_{p=1}^{P} a_p^k \cdot \widehat{\omega}_{i,k,p} \cdot \delta(\omega - \omega_p).
\end{equation}
The global loss function is defined as the weighted average of all local losses:
\begin{equation}
\mathcal{L}(\boldsymbol{\lambda}) := \frac{1}{\sum_{n=1}^{N} |\mathcal{D}_n|} \sum_{n=1}^{N} |\mathcal{D}_n| \, \mathcal{L}_n(\boldsymbol{\lambda}).
\end{equation}
The federated learning objective is to determine the optimal global parameters:
\begin{equation}
\boldsymbol{\lambda}^* = \arg \min_{\boldsymbol{\lambda}} \mathcal{L}(\boldsymbol{\lambda}).
\end{equation}
During each training iteration $t$, local model parameters are updated using the Adam optimizer as
\begin{equation}
\boldsymbol{\lambda}_n^{t+1} = \boldsymbol{\lambda}_n^t - \eta \nabla \mathcal{L}_n(\boldsymbol{\lambda}_n^t),
\end{equation}
where $\eta$ is the learning rate. The global parameters are then updated by aggregating the local updates:
\begin{equation}
\boldsymbol{\lambda}^{t+1} = \sum_{n=1}^{N} \frac{|\mathcal{D}_n|}{\sum_{n=1}^{N} |\mathcal{D}_n|} \boldsymbol{\lambda}_n^{t+1}.
\end{equation}

The RL agent is trained over 2,048 iterations with a batch size of 1,024 and a learning rate of $\eta = 10^{-3}$. The number of agents participating in the federated setting is set to 10. During training, the number of particles $P$ used to represent the posterior distribution is chosen according to the maximum frequency range: $P = 240$ for $\omega_{\text{max}} = 5~\mathrm{MHz}$ and $10~\mathrm{MHz}$, and $P = 240\times2$ for $\omega_{\text{max}} = 20~\mathrm{MHz}$. For performance evaluation in the online inference phase, 2,000 test episodes are conducted. The coherence time of the Ramsey experiment $T_{2}$ is set as $96 \mu$s~\cite{bonato2016optimized}.

The number of local agents \(N\) in our federated RL framework was chosen based on a trade-off between learning performance and computational cost. Specifically, we selected \(N = 10\) to divide the global estimation range \([0, \omega_{\max}]\) into 10 equal subranges, enabling each agent to focus on a simpler local task. This modularization improves policy generalization and convergence speed, compared to training a single global agent over the full range.

From a learning accuracy perspective, training a single agent over the full \(\omega\)-range leads to a significantly harder optimization problem due to the broader and more complex parameter space. This often results in slow convergence and poor generalization. In contrast, decomposing the task into local agents over smaller subranges improves convergence speed, policy generalization within each subrange, and global performance after aggregation via FedAvg. From a learning efficiency perspective, the federated setup also has implications for training cost. We compare it to a single-agent baseline in terms of memory and time complexity in parallel:
	\begin{center}
		\begin{tabular}{|c|c|c|}
			\hline
			& \textbf{Baseline RL} & \textbf{Federated RL} \\
			\hline
			Memory Complexity & \( \mathcal{O}(|\mathcal{D}_{n}| \cdot M \cdot P) \) & \( \mathcal{O}(N \cdot |\mathcal{D}_{n}| \cdot M \cdot P) \) \\
			Time Complexity & \( \mathcal{O}(M) \) & \( \mathcal{O}(M) \) \\
			\hline
		\end{tabular}
	\end{center}
	To assess the sensitivity of our method to the number of agents, we conducted an ablation study with \(N \in \{5,\, 10,\, 15\}\). To ensure consistent particle resolution per agent, the corresponding number of particles was set as \(P \in \{ 240 \times 2,\, 240,\, 240\}\), respectively.
 The results in Supplementary Figure~\ref{fig:5} illustrate the impact of \( N \) on the training performance and the final estimation mean squared error (MSE). We observe that training performance improves with increasing \( N \), indicating that too few agents \( N = 5 \) limit the benefits of task decomposition and make local learning less effective. Conversely, too many agents (\( N = 15 \)) increase memory or time costs without significant gains in accuracy.
    
	\begin{figure}[H]
		\centering
		\begin{subfigure}[t]{0.49\textwidth}
			\centering
			\includegraphics[width=\linewidth]{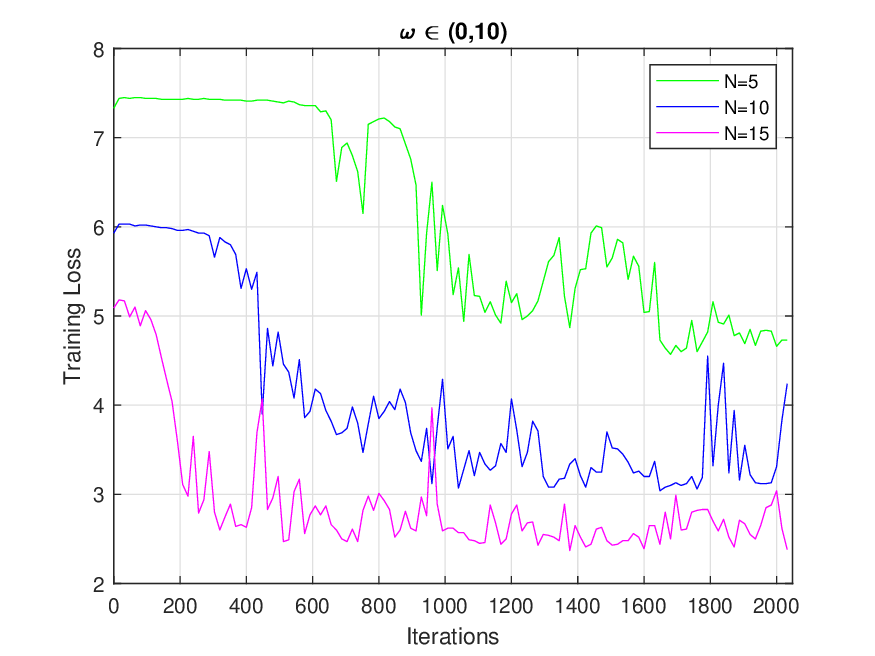}
			\caption{Federated-RL training loss}
			\label{fig:5a}
		\end{subfigure}
		\hfill
		\begin{subfigure}[t]{0.49\textwidth}
			\centering
			\includegraphics[width=\linewidth]{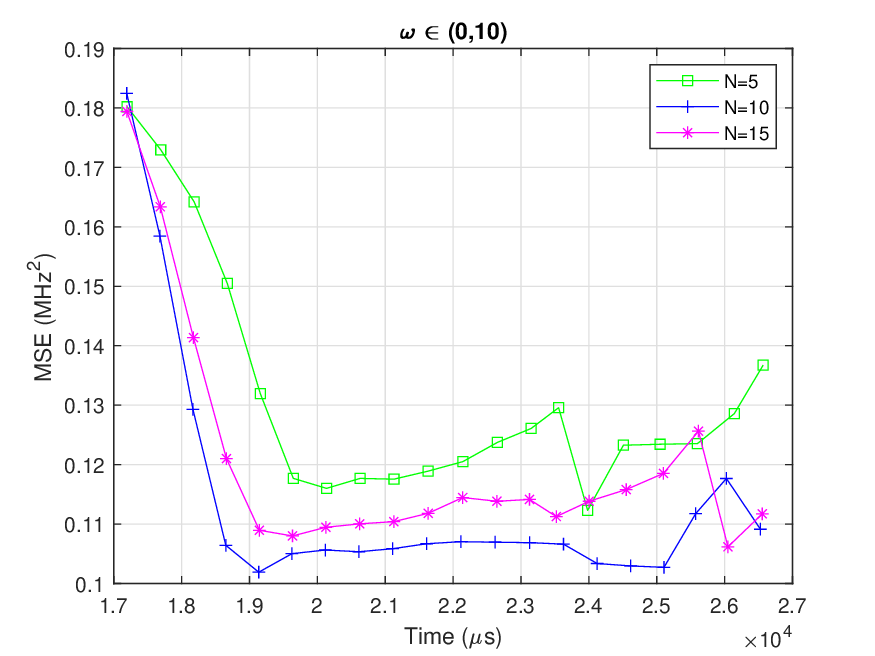}
			\caption{End-to-end performance}
			\label{fig:5c}
		\end{subfigure}
        \captionsetup{name=Supplementary Figure}
		\caption{Training and evaluation performance as affected by the number of agents.}
		\label{fig:5}
\end{figure}

\section{Complexity Analysis of the Protocol}\label{secMethods-complexity}
The computational overhead associated with both the BNN and the federated RL components is summarized in the insights below.

\subsection{Offline Training Phase – BNN}

\textit{Training Time Complexity.} Assuming \( |\mathcal{D}_{\omega}| \) weight samples per input and a batch size \( b_{\omega} \), the BNN training time per step scales as:
	\[
	\text{Time}_{\text{BNN-Train}} = \mathcal{O}(|\mathcal{D}_{\omega}| \cdot b_{\omega} \cdot S_{1}),
	\]
	where \( S_{1} \) is the number of trainable parameters in the BNN, typically a small multilayer perceptron with 32 neurons per layer.
    
	\textit{Memory Complexity.} The memory overhead primarily comes from intermediate activations and gradient storage over all samples:
	\[
	\text{Memory}_{\text{BNN-Train}} = \mathcal{O}(|\mathcal{D}_{\omega}| \cdot b_{\omega} \cdot 32 + S_{1}).
	\]
    
	\textit{Inference Cost.} During deployment, we use \( |\mathcal{D}_{\omega}|_{\text{test}} \in [10, 100] \) Monte Carlo samples for posterior averaging:
	\[
	\text{Time}_{\text{BNN-Infer}} = \mathcal{O}(|\mathcal{D}_{\omega}|_{\text{test}} \cdot b_{\omega} \cdot S_{1}), \quad
	\text{Memory}_{\text{BNN-Infer}} = \mathcal{O}(|\mathcal{D}_{\omega}|_{\text{test}} \cdot b_{\omega} \cdot 32 + S_{1}).
	\]
	Inference is substantially cheaper than training and can be accelerated by using fewer samples or mean weights.
    
\subsection{Offline Training Phase – Federated RL}
    
	\textit{Memory Complexity.} For each RL agent trained over \( |\mathcal{D}_{n}| \) samples and \( M \) measurements with \( P \) particles:
	\[
	\text{Memory}_{\text{per agent}} = \mathcal{O}(|\mathcal{D}_{n}| \cdot M \cdot P).
	\]
	The total memory scales as:
	\[
	\text{Memory}_{\text{FedRL}} =
	\begin{cases}
		\mathcal{O}(  M ), & \text{(sequential)}\\
		\mathcal{O}(N \cdot |\mathcal{D}_{n}| \cdot M \cdot P), & \text{(parallel) }
	\end{cases}
	\]
    
	\textit{Time Complexity.} The time cost per update step per agent is:
	\[
	\text{Time}_{\text{FedRL}} =
	\begin{cases}
		\mathcal{O}( M ), & \text{(parallel)}\\
		\mathcal{O}(N \cdot |\mathcal{D}_{n}| \cdot M \cdot P), & \text{(sequential)}
	\end{cases}
	\]
	FedAvg incurs only a minor overhead of \( \mathcal{O}(S_{2}) \), where \( S_{2} \) is the RL model size.
    
\subsection{Inference Phase – BNN + Federated RL Agent}
    
	At runtime, only the global RL policy and BNN are used for adaptive control. The per-measurement inference cost is:
	\[
	\text{Time}_{\text{Inference}} = \mathcal{O}(S_{1}+M),
	\]
	which matches the cost of the single-agent baseline. Local agents are not involved at test time.
    
In our simulation experimental setup, the total inference time per measurement (BNN + Federated RL) is approximately \( 70 \sim 140 \, \mu\mathrm{s} \), comparable to the NV spin initialization time (\( \sim 200 \, \mu\mathrm{s} \))~\cite{bonato2016optimized}. Moreover, the initialization and inference stages can be executed in parallel, enabling real-time sensing without additional latency. Although BNN training incurs extra cost due to sampling, this is efficiently handled by GPU parallelism. The federated RL module scales with agent count in memory, but can be parallelized for efficient training. Most importantly, inference cost is low and equivalent to baseline methods, ensuring practical real-time deployment. All simulations in this work were performed on an NVIDIA RTX 4090 GPU.

\section{Robustness Analysis of the Protocol}\label{secResult-robust}
For the realistic noisy scenario, we model asymmetric readout fidelities as $F_1 = 0.993$ and $F_0 = C + (1 - F_1)$, where the experimentally achievable contrast is set to $C = 0.99$. These values reflect performance typically attainable in room-temperature NV center experiments with $50{,}000$ repetitions, thereby ensuring our simulations closely mirror practical sensing conditions.

We simulate 1024 independent test trajectories for each configuration over 200 iterations, yielding $1024 \times 200$ samples per condition. The statistical error is quantified using the 95$\%$ confidence interval (CI95), computed as: $\text{CI95} = \bar{x} \pm t_{0.975, d-1} \cdot \frac{i}{\sqrt{d}},$ where $\bar{x}$ is the sample mean, $i$ is the sample standard deviation, and $t_{0.975, d-1}$ is the critical value from the t-distribution with $d - 1$ degrees of freedom.
    
\begin{figure}[H]
		\centering
		\begin{subfigure}[t]{0.49\textwidth}
			\centering
			\includegraphics[width=\linewidth]{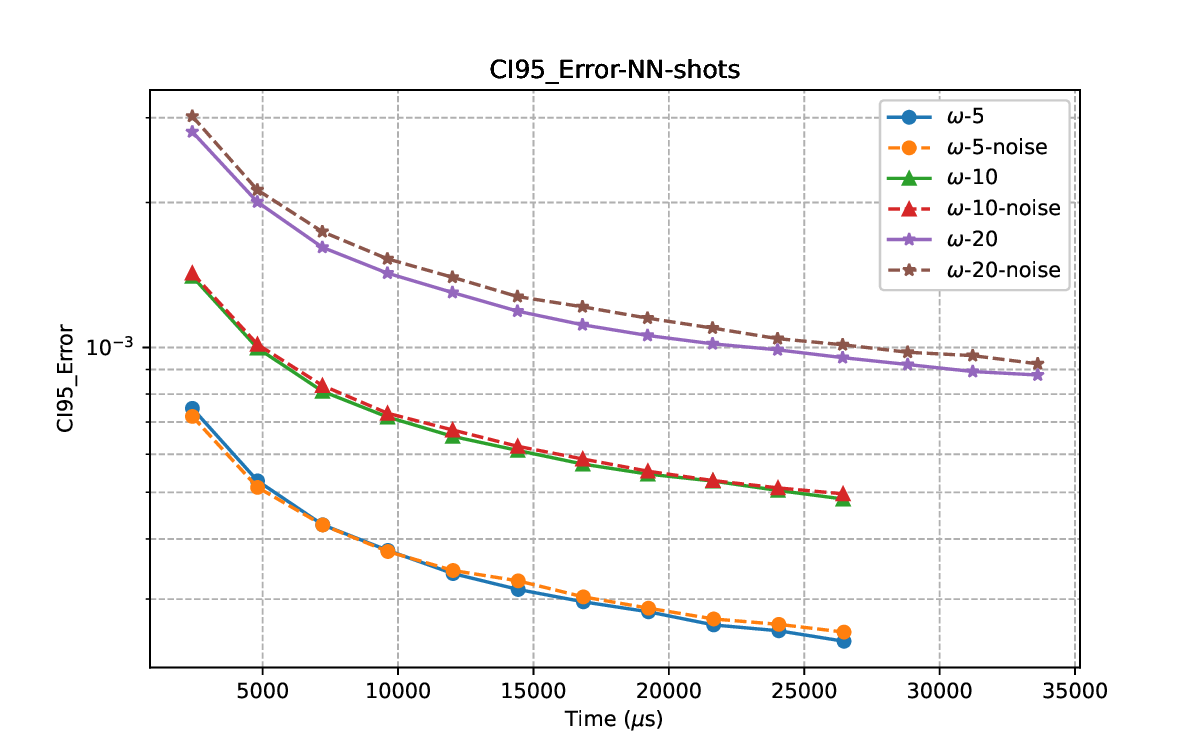}
			\caption{1$^{\text{st}}$-stage BNN}
			\label{fig:3a}
		\end{subfigure}
		\hfill
		\begin{subfigure}[t]{0.49\textwidth}
			\centering
			\includegraphics[width=\linewidth]{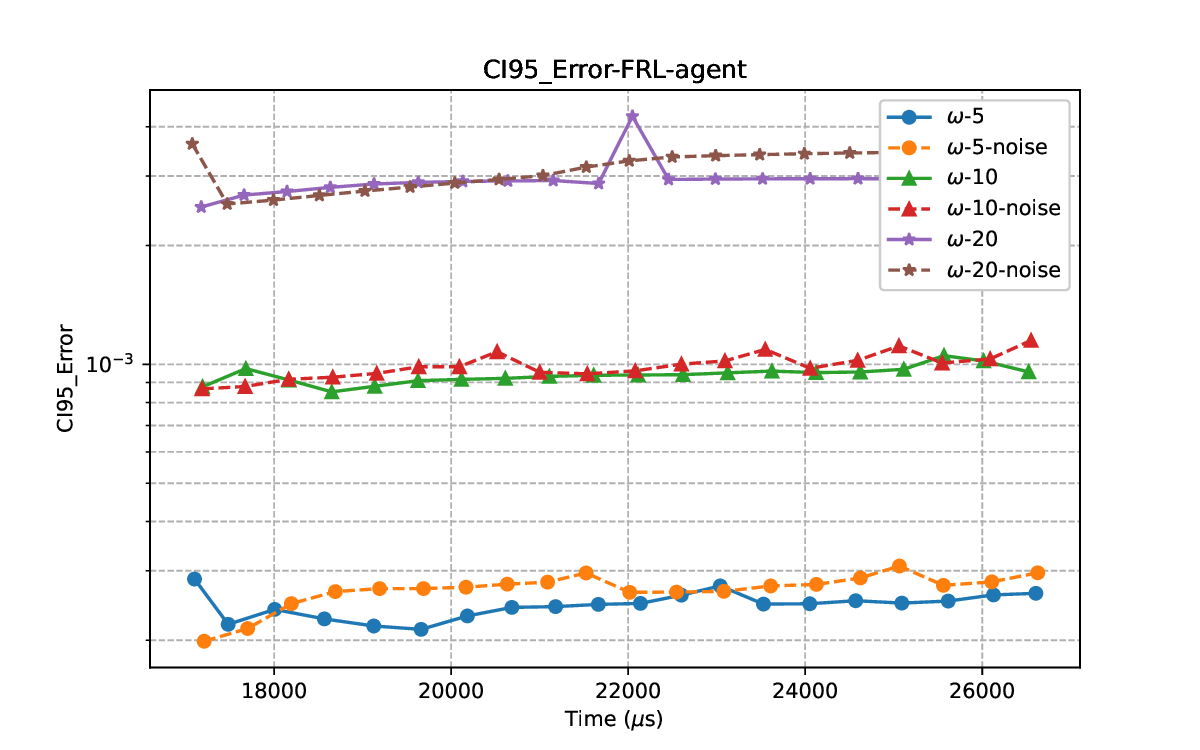}
			\caption{2$^{\text{nd}}$-stage federated RL agent}
			\label{fig:3b}
		\end{subfigure}
        \captionsetup{name=Supplementary Figure}
		\caption{Statistical error of results for the 1$^{\text{st}}$-stage BNN and 2$^{\text{nd}}$-stage federated RL agent.}
		\label{fig:3}
	\end{figure}

Below, we summarize the CI95 errors for both noise-free and noisy conditions under various $\omega_{\max}$ values, including the 1$^{\text{st}}$-stage BNN and 2$^{\text{nd}}$-stage adaptive estimation performance. Supplementary Figure~\ref{fig:3} shows the statistical error of BNN and federated RL agent under testing by $1,024 \times 200$ samples. Supplementary Figure~\ref{fig:3}(a) shows the statistical error of BNN after 10--110 measurement shots under $\omega \in (0, 5)$ MHz and $\omega \in (0, 10)$ MHz, and after 10--140 shots under $\omega \in (0, 20)$ MHz. Supplementary Figure~\ref{fig:3}(b) shows the statistical error of the federated RL agent when the total time budget is 10 ms. To specify the statistical error of BNN after the 1$^{\text{st}}$-stage (i.e., after 70 shots in this work), the table lists CI95 values. The CI95 for the 2$^{\text{nd}}$-stage is the mean CI95 over the whole measurement time budget.

	\begin{table}[h]
		\centering
        \captionsetup{name=Supplementary Table}
		\caption{CI95 results for different ranges of $\omega$.}
		\begin{tabular}{|c|c|c|}
			\hline
			Ranges of $\omega$ (MHz) & CI$_{95}$ (with noise / w/o noise) & bandwidth-normalized CI$_{95}$ \\
			\hline
			1st Stage: $\omega_{\max} = 5$       & 0.0296\% / 0.0303\%         & $\sim$0.0059\% / 0.0061\% \\
			1st Stage: $\omega_{\max} = 10$      & 0.0572\% / 0.0587\%         & $\sim$0.0057\% / 0.0059\% \\
			1st Stage: $\omega_{\max} = 20$      & 0.1115\% / 0.1215\%         & $\sim$0.0056\% / 0.0061\% \\
			1st Stage (100 shots): $\omega_{\max} = 20$ & 0.0989\% / 0.1043\%     & $\sim$0.0049\% / 0.0052\% \\
			\hline
			2nd Stage: $\omega_{\max} = 5$       & 0.02458\% / 0.02695\%       & 0.00492\% / 0.00539\% \\
			2nd Stage: $\omega_{\max} = 10$      & 0.09409\% / 0.09934\%       & 0.00941\% / 0.00993\% \\
			2nd Stage: $\omega_{\max} = 20$      & 0.29998\% / 0.31687\%       & 0.01500\% / 0.01584\% \\
			\hline
		\end{tabular}
		\label{table:1}
	\end{table}
	\begin{figure}[htbp]
		\centering
		\begin{tikzpicture}
			\node[inner sep=0pt] (grid) at (0,0) {
				\begin{minipage}{\textwidth}
					\centering
					\begin{tabular}{ccc}
						\includegraphics[width=0.31\textwidth]{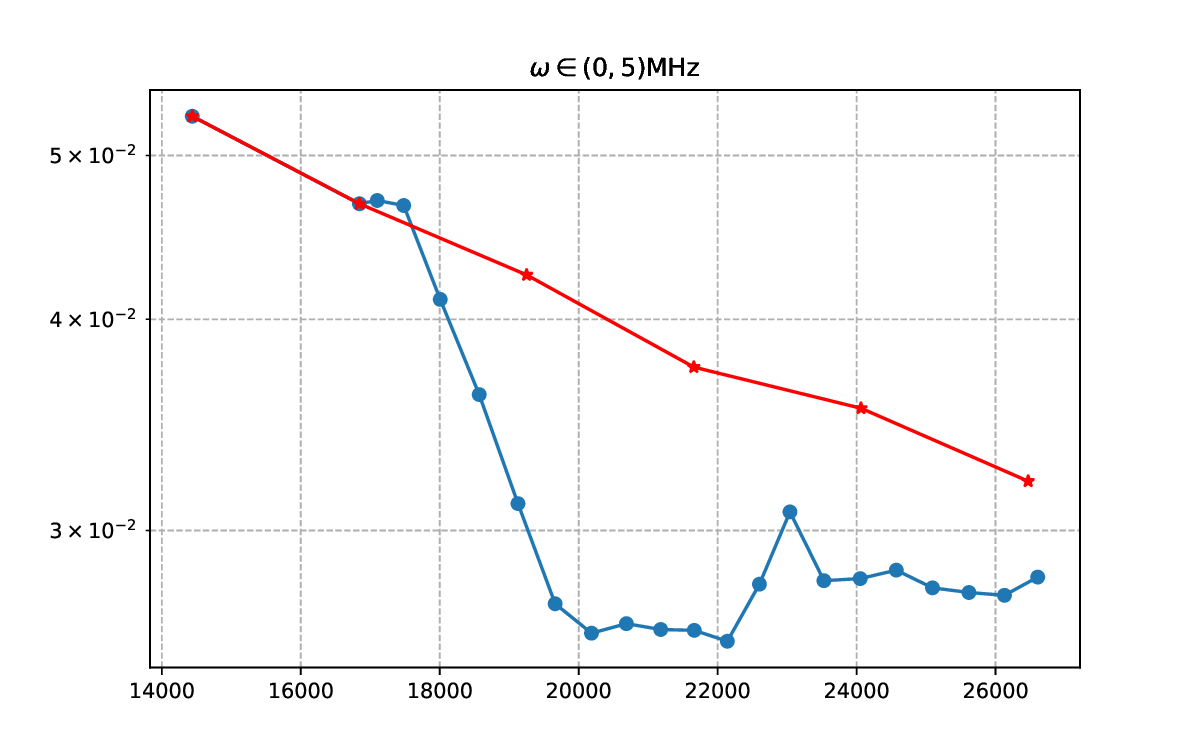} &
						\includegraphics[width=0.31\textwidth]{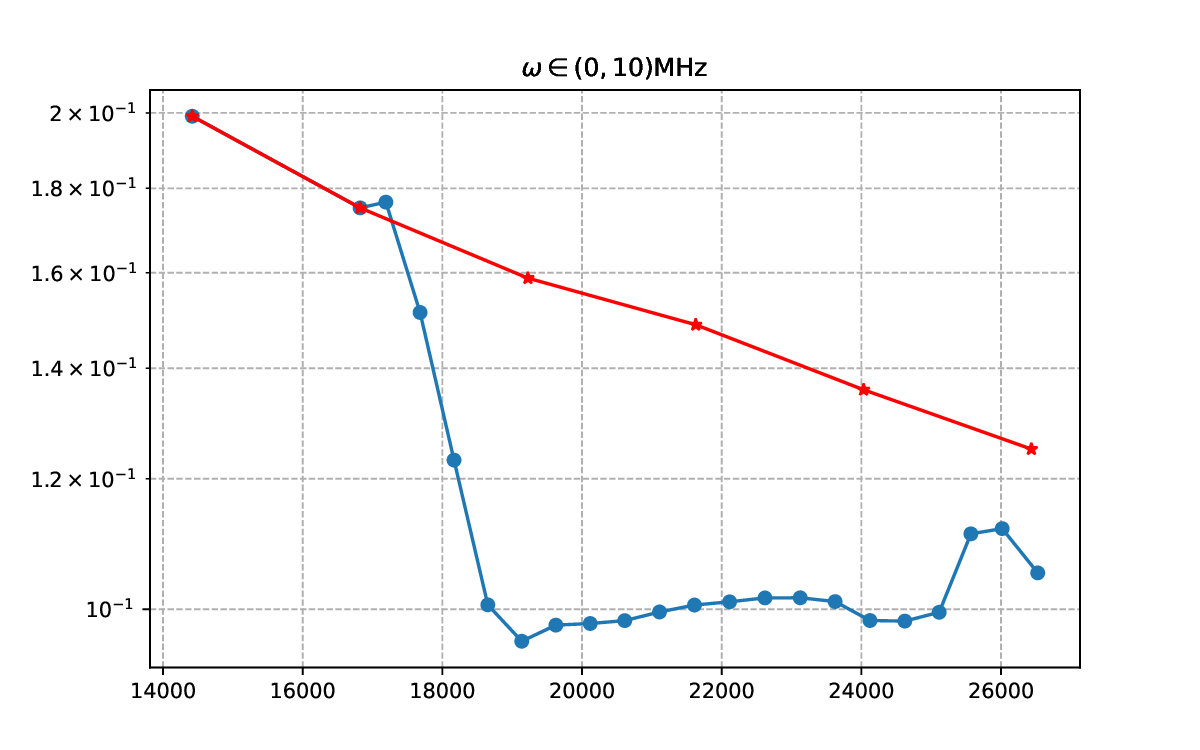} &
						\includegraphics[width=0.31\textwidth]{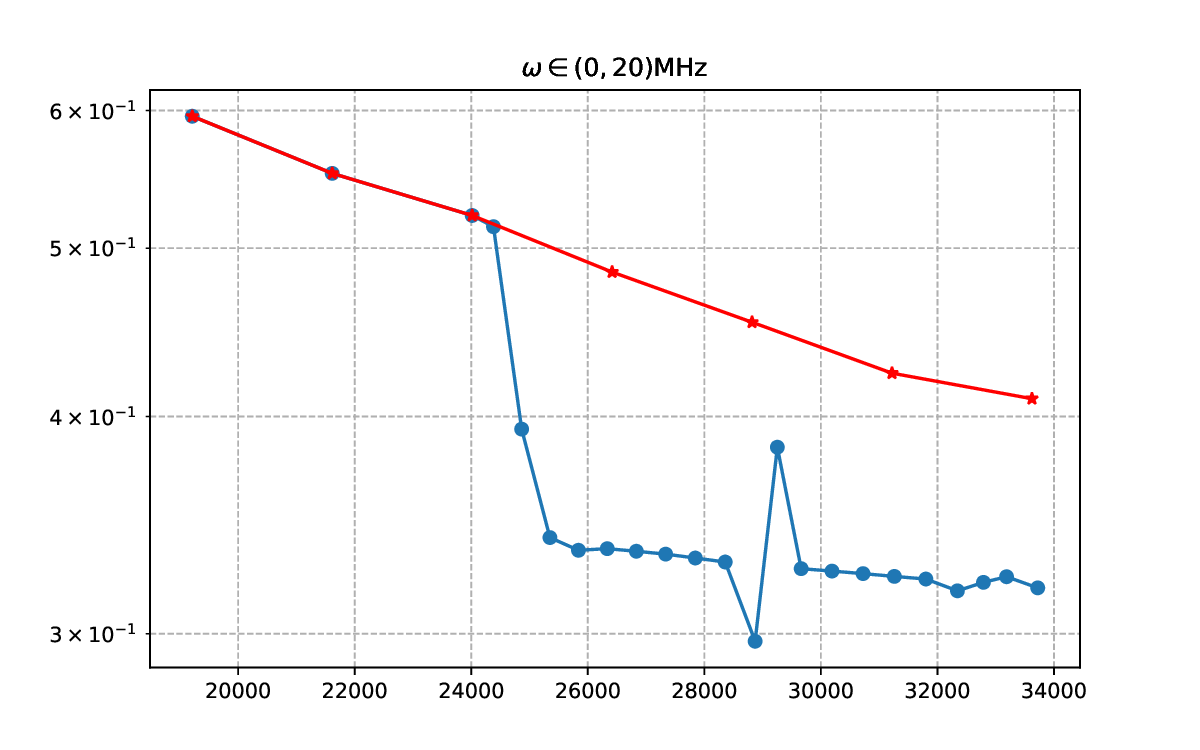} \\
						\includegraphics[width=0.31\textwidth]{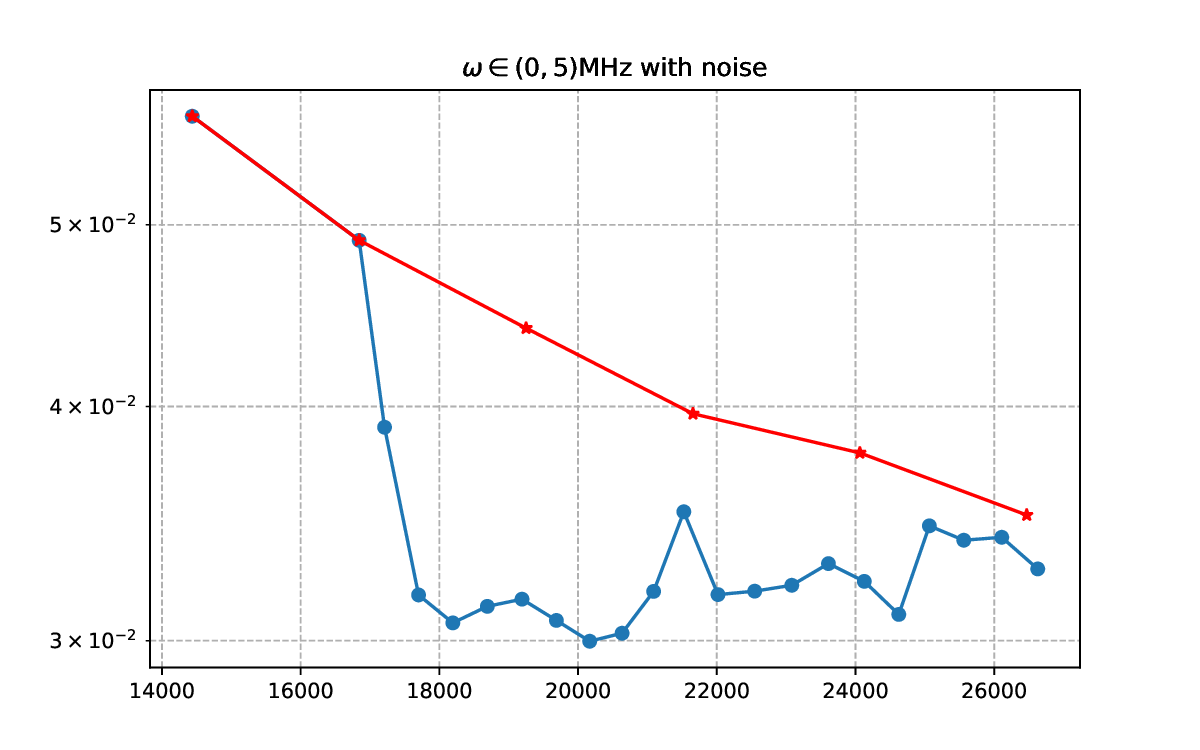} &
						\includegraphics[width=0.31\textwidth]{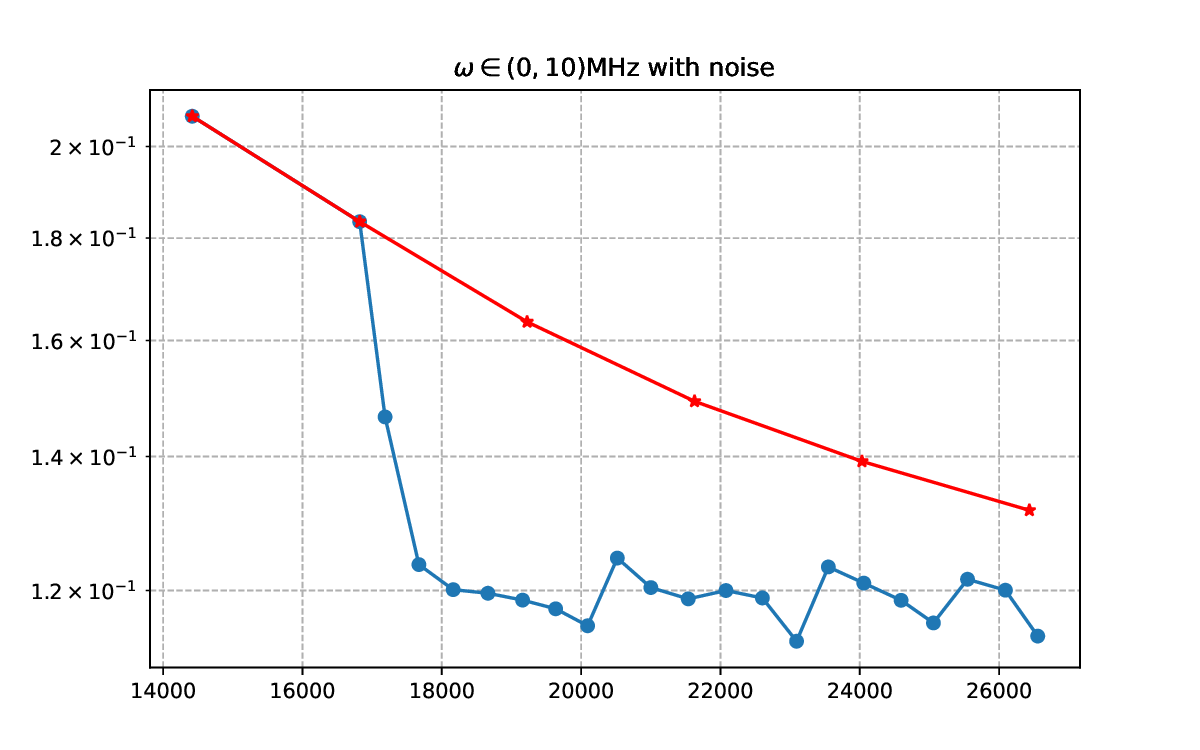} &
						\includegraphics[width=0.31\textwidth]{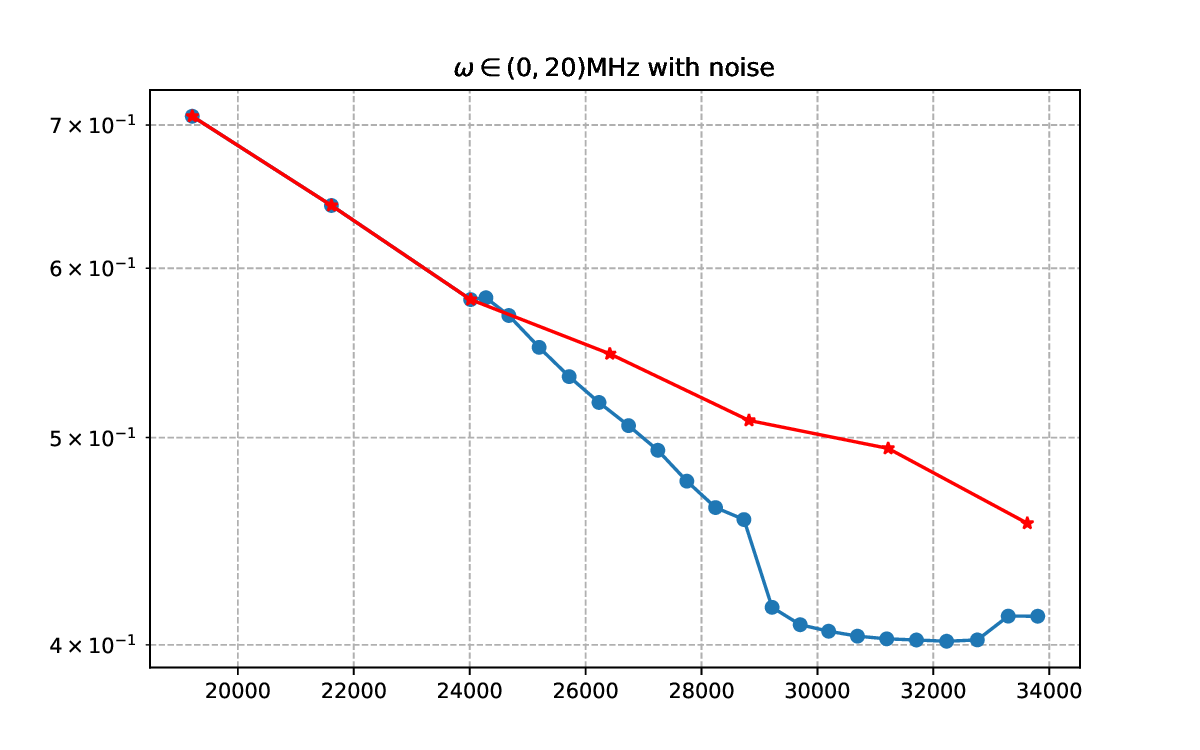} \\
					\end{tabular}
				\end{minipage}
			};
			\node[yshift=-3.3cm]{Time ($\mu$s)};
			\node[rotate=90] at ([xshift=0cm]grid.west) {MSE (MHz${}^{2}$)};
		\end{tikzpicture}
        \captionsetup{name=Supplementary Figure}
		\caption{Comparison of the proposed scheme with NN-shots under noisy and noise-free conditions.}
		\label{fig:4}
	\end{figure}

    In Supplementary Figure~\ref{fig:4}, the red curve represents the NN-shots approach, while the blue curve corresponds to the performance of the proposed method. The 1$^{\text{st}}$ Stage concludes after 70 measurement shots for the cases $\omega \in (0, 5)$~MHz and $\omega \in (0, 10)$~MHz. For the case $\omega \in (0, 20)$~MHz, the 1$^{\text{st}}$ Stage ends after 100 measurement shots, and the time budget for the 2$^{\text{nd}}$ Stage is set to 10~ms. As shown in Supplementary Figure~\ref{fig:4}, under both noise-free and noisy conditions, the proposed scheme consistently outperforms the NN-shots strategy within the given resource constraints. Moreover, as shown in Supplementary Table~\ref{table:1}, the CI$_{95}$ intervals are consistently tight, even in noisy settings and across a wide range of $\omega_{\max}$. Therefore, the results in Supplementary Table~\ref{table:1} and Supplementary Figure~\ref{fig:4} collectively demonstrate the stability and reliability of the simulation results and support the practical feasibility of the proposed protocol in realistic experimental environments.
     
\section{Runtime Evolution of Sensing Parameters}\label{secResult-evolution}
\begin{figure}[htbp]
	\centering
	\begin{subfigure}[b]{0.31\textwidth}
		\includegraphics[width=\linewidth]{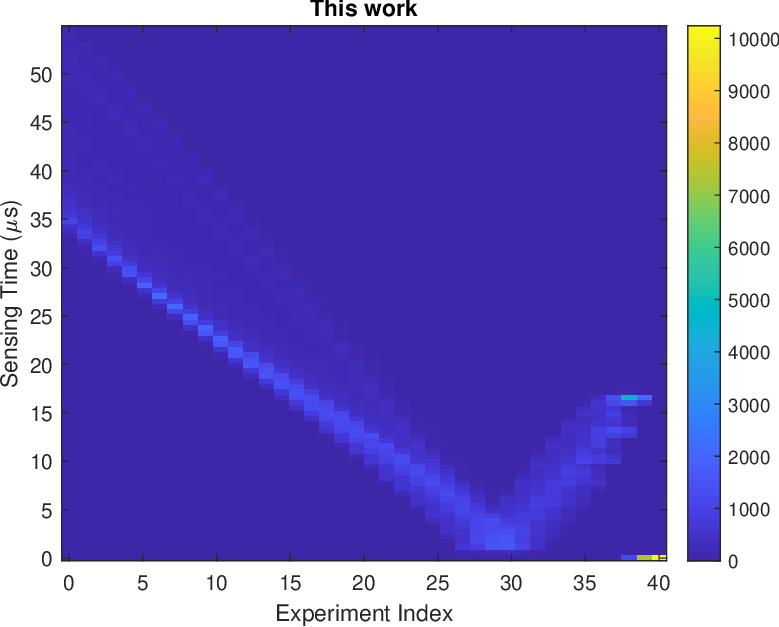}
		\caption*{}
	\end{subfigure}
	\begin{subfigure}[b]{0.31\textwidth}
		\includegraphics[width=\linewidth]{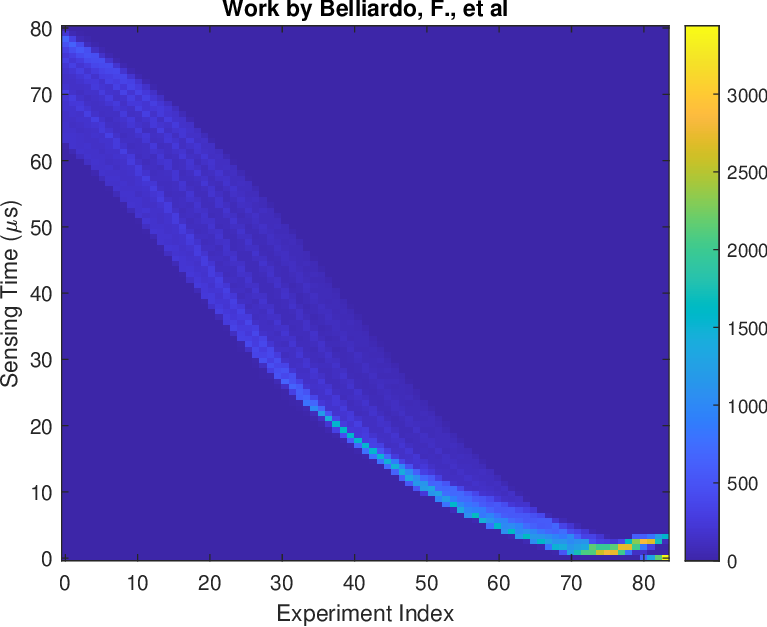}
		\caption*{$\omega \in (0,5)$MHz }
	\end{subfigure}
	\begin{subfigure}[b]{0.31\textwidth}
		\includegraphics[width=\linewidth]{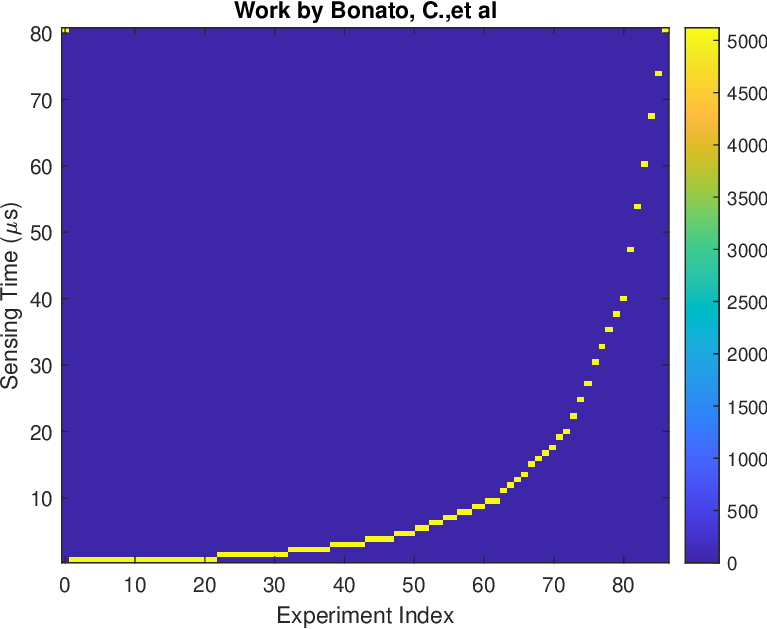}
		\caption*{}
	\end{subfigure}
	
	\vspace{5pt} 
	
	\begin{subfigure}[b]{0.31\textwidth}
		\includegraphics[width=\linewidth]{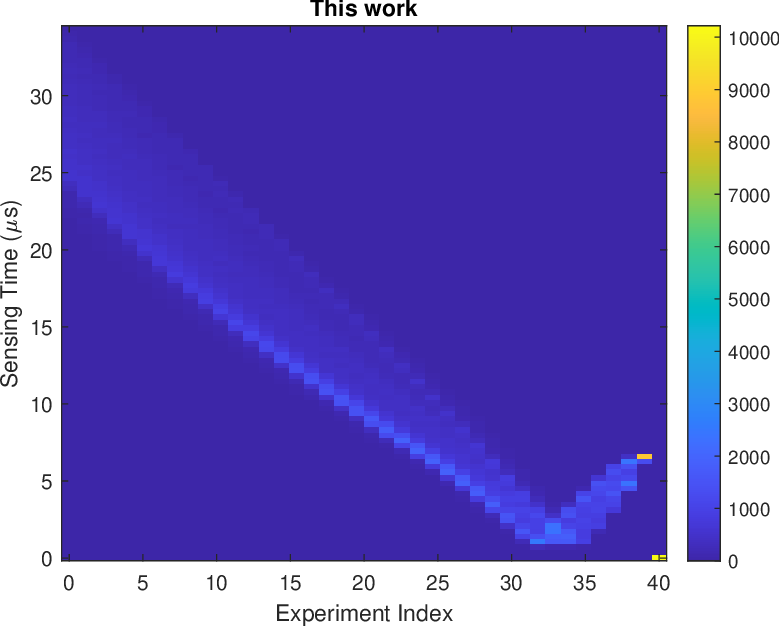}
		\caption*{}
	\end{subfigure}
	\begin{subfigure}[b]{0.31\textwidth}
		\includegraphics[width=\linewidth]{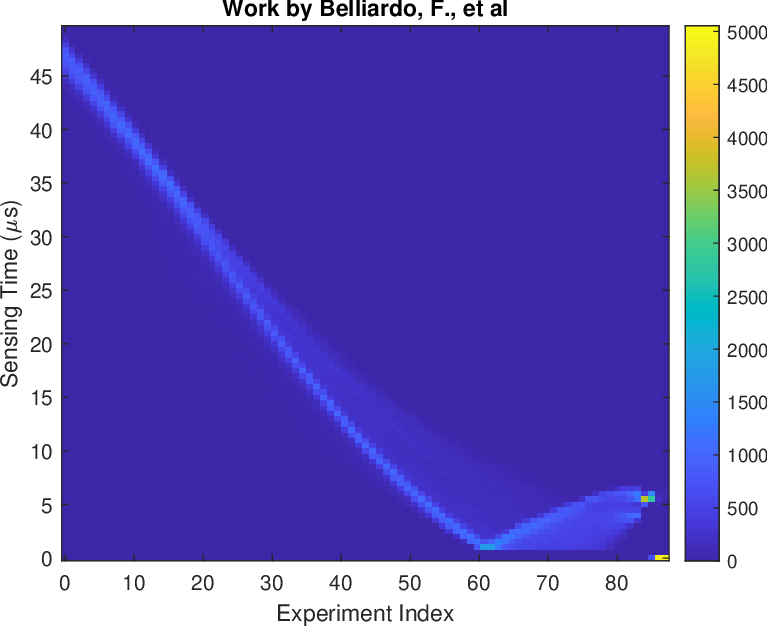}
		\caption*{$\omega \in (0,10)$MHz }
	\end{subfigure}
	\begin{subfigure}[b]{0.31\textwidth}
		\includegraphics[width=\linewidth]{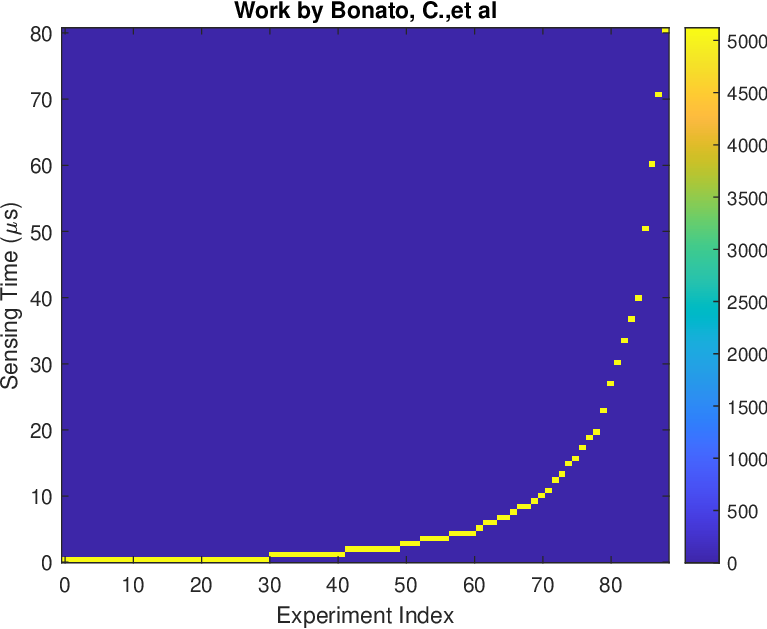}
		\caption*{}
	\end{subfigure}
	
	\vspace{5pt} 
	
	\begin{subfigure}[b]{0.31\textwidth}
		\includegraphics[width=\linewidth]{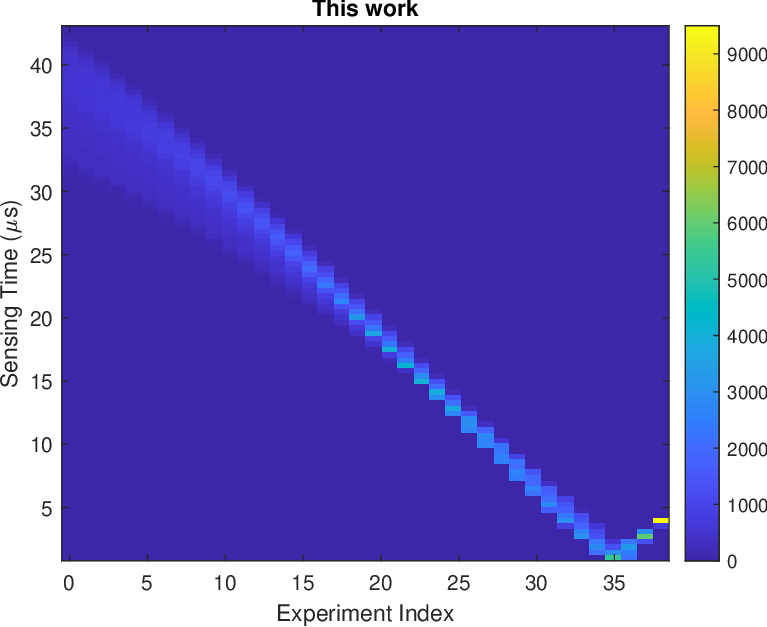}
		\caption*{}
	\end{subfigure}
	\begin{subfigure}[b]{0.31\textwidth}
		\includegraphics[width=\linewidth]{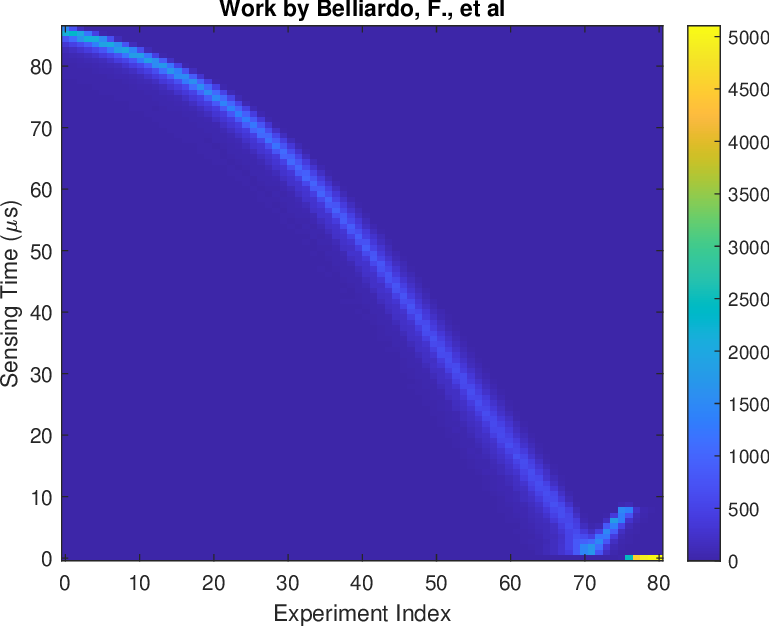}
		\caption*{$\omega \in (0,20)$MHz }
	\end{subfigure}
	\begin{subfigure}[b]{0.31\textwidth}
		\includegraphics[width=\linewidth]{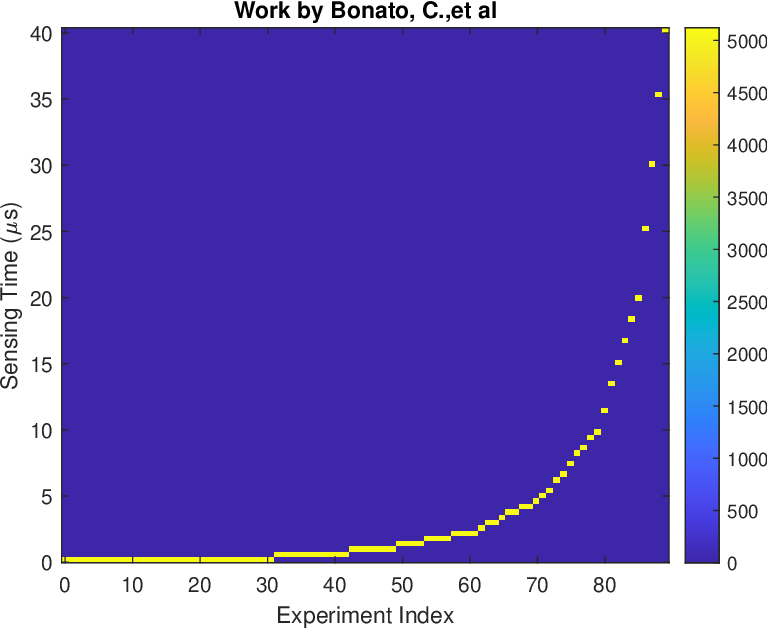}
		\caption*{}
	\end{subfigure}
	\captionsetup{name=Supplementary Figure}
	\caption{Distribution of optimized sensing time for different protocols}
	\label{fig:tau}
\end{figure}

\begin{figure}[htbp]
	\centering
	\begin{subfigure}[b]{0.31\textwidth}
		\includegraphics[width=\linewidth]{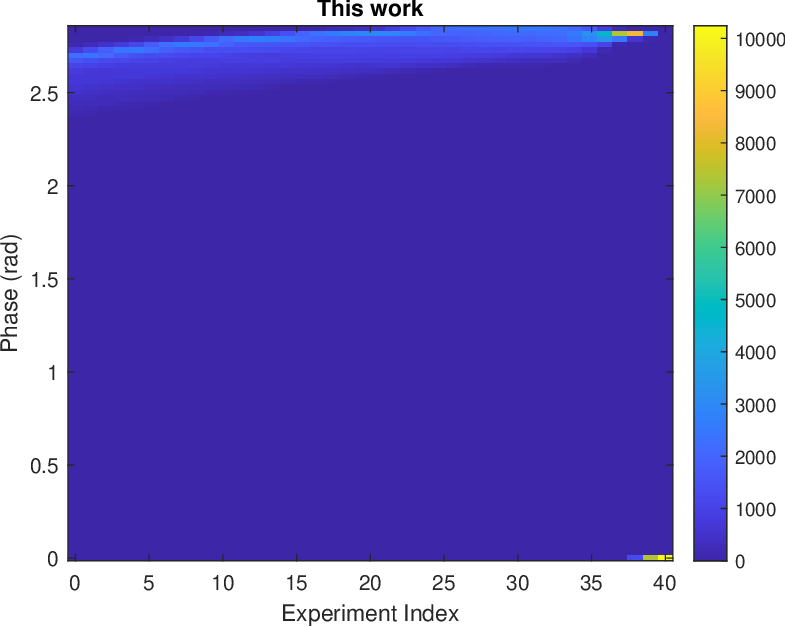}
		\caption*{}
	\end{subfigure}
	\begin{subfigure}[b]{0.31\textwidth}
		\includegraphics[width=\linewidth]{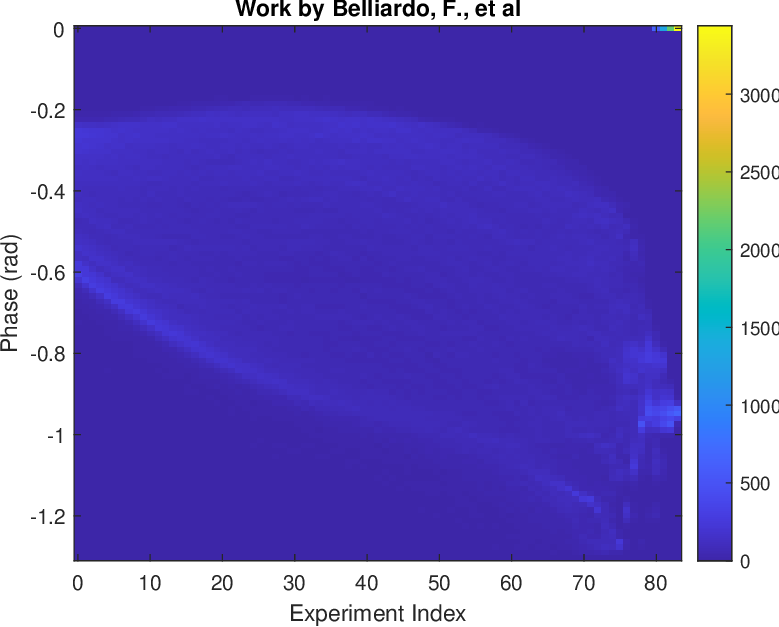}
		\caption*{$\omega \in (0,5)$MHz }
	\end{subfigure}
	\begin{subfigure}[b]{0.31\textwidth}
		\includegraphics[width=\linewidth]{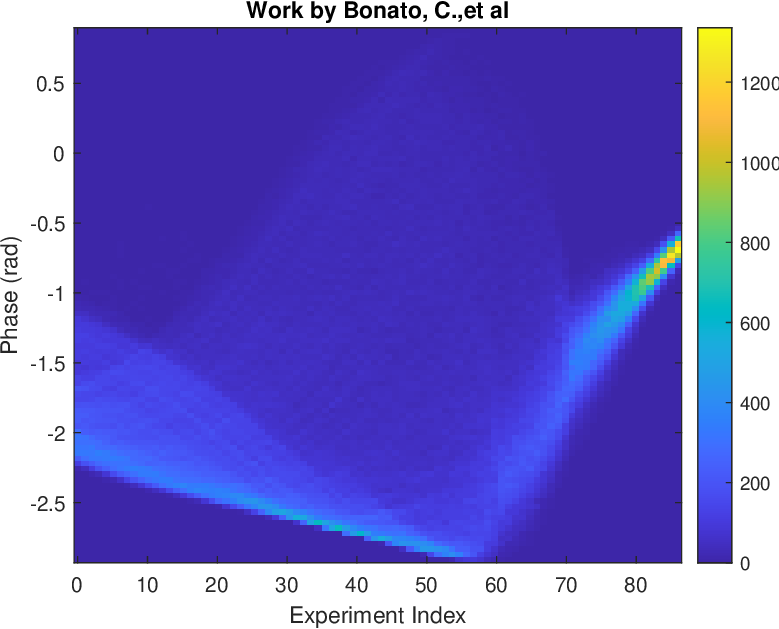}
		\caption*{}
	\end{subfigure}
	
	\vspace{5pt} 
	
	\begin{subfigure}[b]{0.31\textwidth}
		\includegraphics[width=\linewidth]{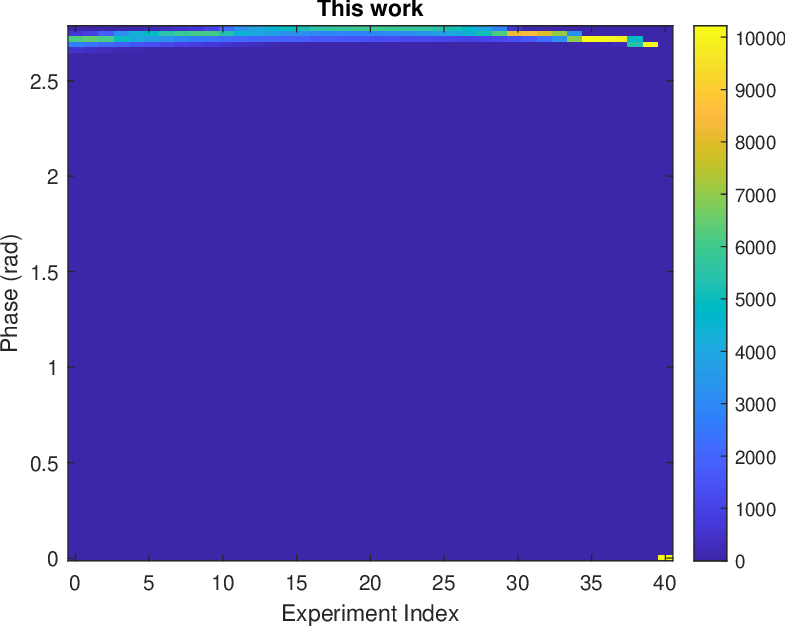}
		\caption*{}
	\end{subfigure}
	\begin{subfigure}[b]{0.31\textwidth}
		\includegraphics[width=\linewidth]{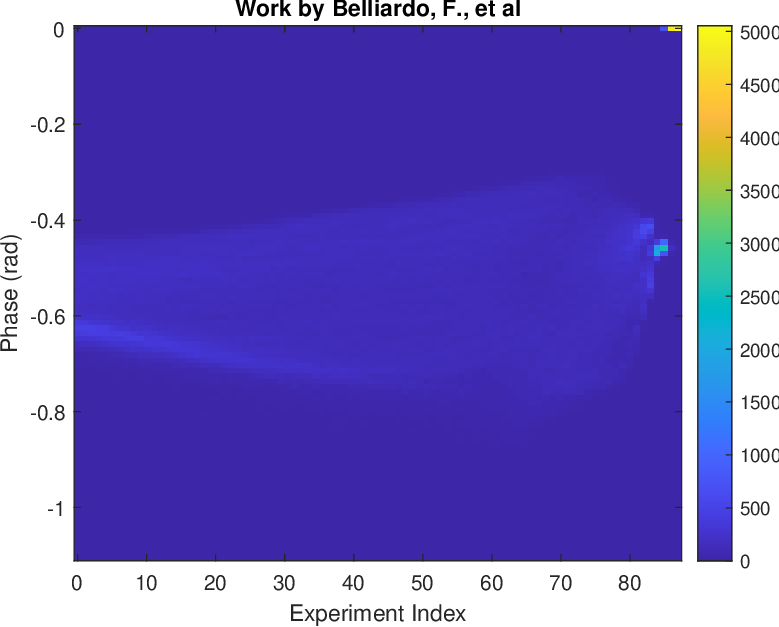}
		\caption*{$\omega \in (0,10)$MHz }
	\end{subfigure}
	\begin{subfigure}[b]{0.31\textwidth}
		\includegraphics[width=\linewidth]{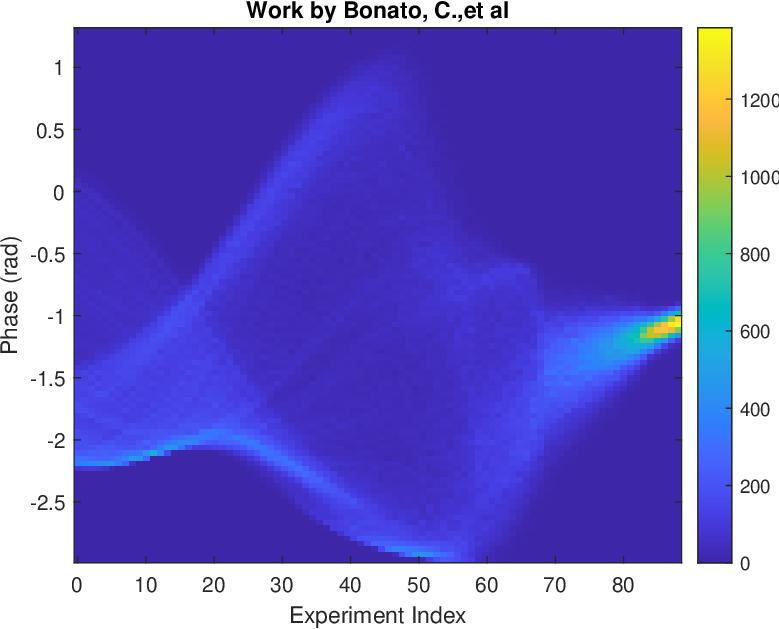}
		\caption*{}
	\end{subfigure}
	
	\vspace{5pt} 
	
	\begin{subfigure}[b]{0.31\textwidth}
		\includegraphics[width=\linewidth]{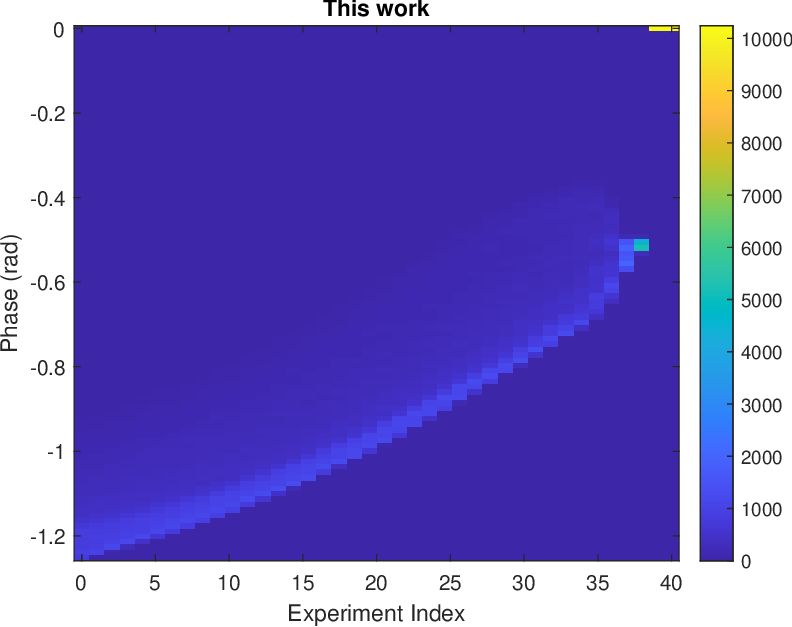}
		\caption*{}
	\end{subfigure}
	\begin{subfigure}[b]{0.31\textwidth}
		\includegraphics[width=\linewidth]{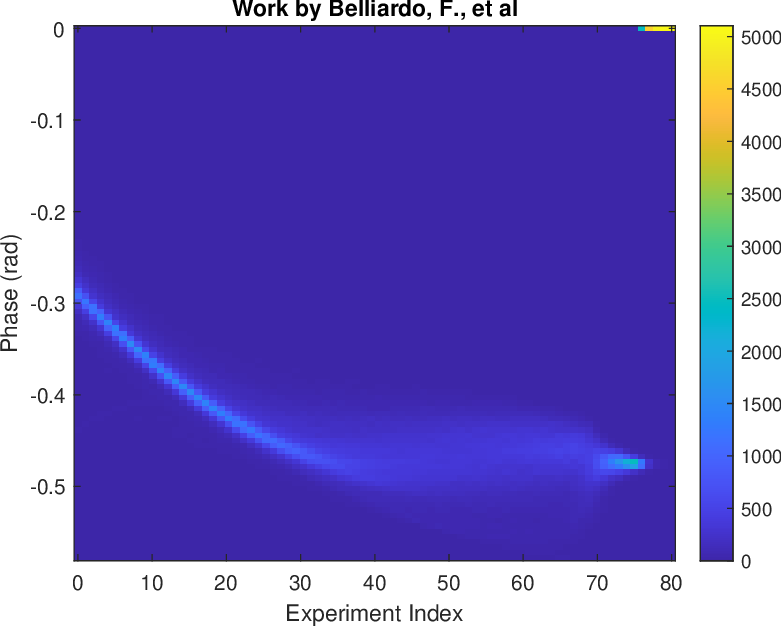}
		\caption*{$\omega \in (0,20)$MHz }
	\end{subfigure}
	\begin{subfigure}[b]{0.31\textwidth}
		\includegraphics[width=\linewidth]{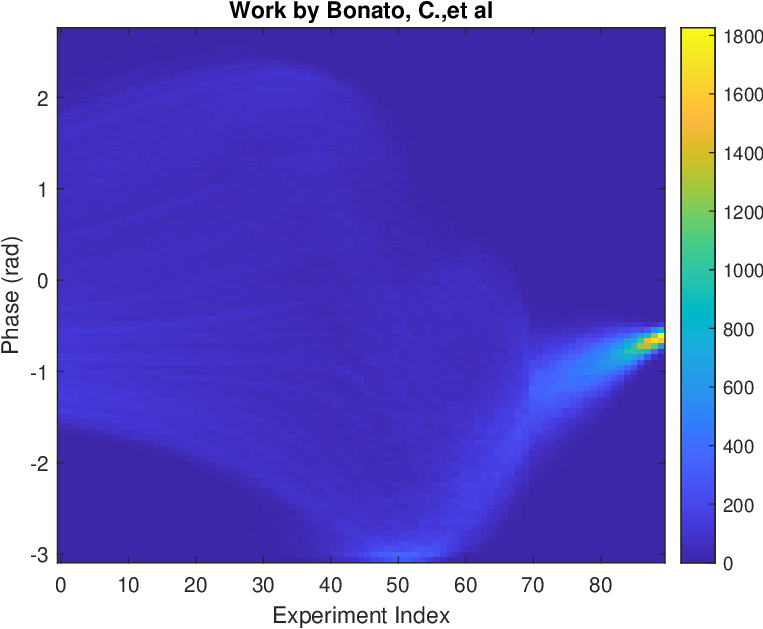}
		\caption*{}
	\end{subfigure}
	\captionsetup{name=Supplementary Figure}
	\caption{Distribution of optimized control phase for different protocols}
	\label{fig:phi}
\end{figure}
Supplementary Figure \ref{fig:tau} and \ref{fig:phi} show the frequency distributions of the optimized sensing parameters that correspond to the estimation results presented in Figure 2 and 3 of the main document. In Supplementary Figure~\ref{fig:tau}, the optimized sensing time of this work's protocol exhibits a trend similar to that reported by Belliardo et al.~\cite{belliardo2024model}. In both cases, the sensing time initially decreases as time resources are dissipated, and then slightly increases near the end of the estimation process as the remaining resources become limited. Nevertheless, the advantage of our protocol is the faster convergence speed as evidenced by the steeper slope and a smaller number of experiments (40 in ours while $>$ 90 in others). This is because the search space of our algorithm is comparably smaller which is in the range of 0-40 $\mu$s, thanks to the narrowed range from the output of the protocol's $1^{\text{st}}$ Stage. It is also worth to mention that the method by Bonato et al. \cite{bonato2016optimized} follows an exponential function to select the sensing time $\tau$. Most of the experiments choose a very small $\tau$ while large values of $\tau$'s are selected very sparsely. This effect is very similar to the method by Nolan et al. \cite{nolan2021machine} in which $\tau$ is fixed as $\tau_{\text{min}}$. This unfortunately results in a significant loss of exploration in the search space. In addition to the sensing time optimization, Supplementary Figure \ref{fig:tau} shows the distribution of controlled phase $\varphi$ during the runs of the three algorithms. It can be observed that the evolution of $\varphi$ has a stark difference among the three protocols, but they all exhibit convergence toward the final stages of the estimation sequence, as evidenced by the bright point on the far right of each figure. The results in Supplementary Figure \ref{fig:phi} further confirm that our protocol converges faster. 

\begin{figure}[htbp]
	\centering
	\begin{subfigure}[b]{0.31\textwidth}
		\includegraphics[width=\linewidth]{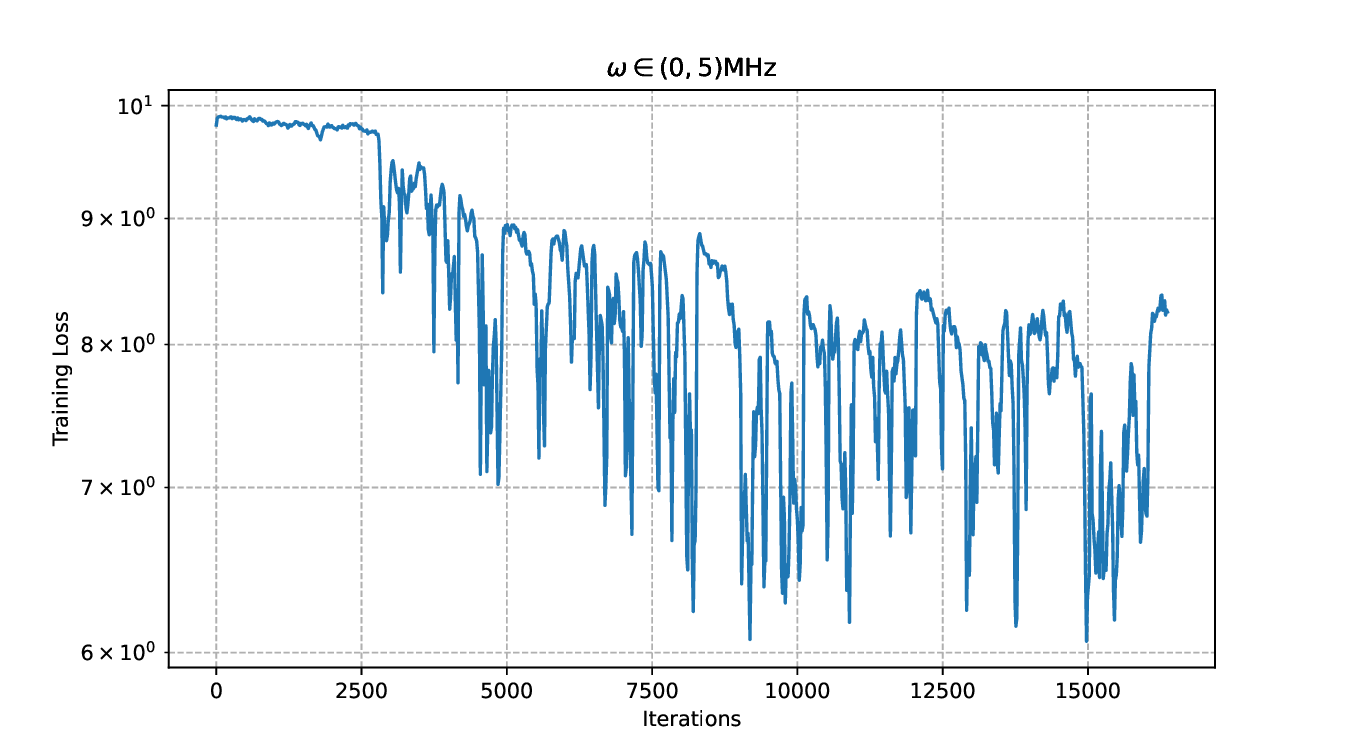}
		\caption*{}
	\end{subfigure}
	\begin{subfigure}[b]{0.31\textwidth}
		\includegraphics[width=\linewidth]{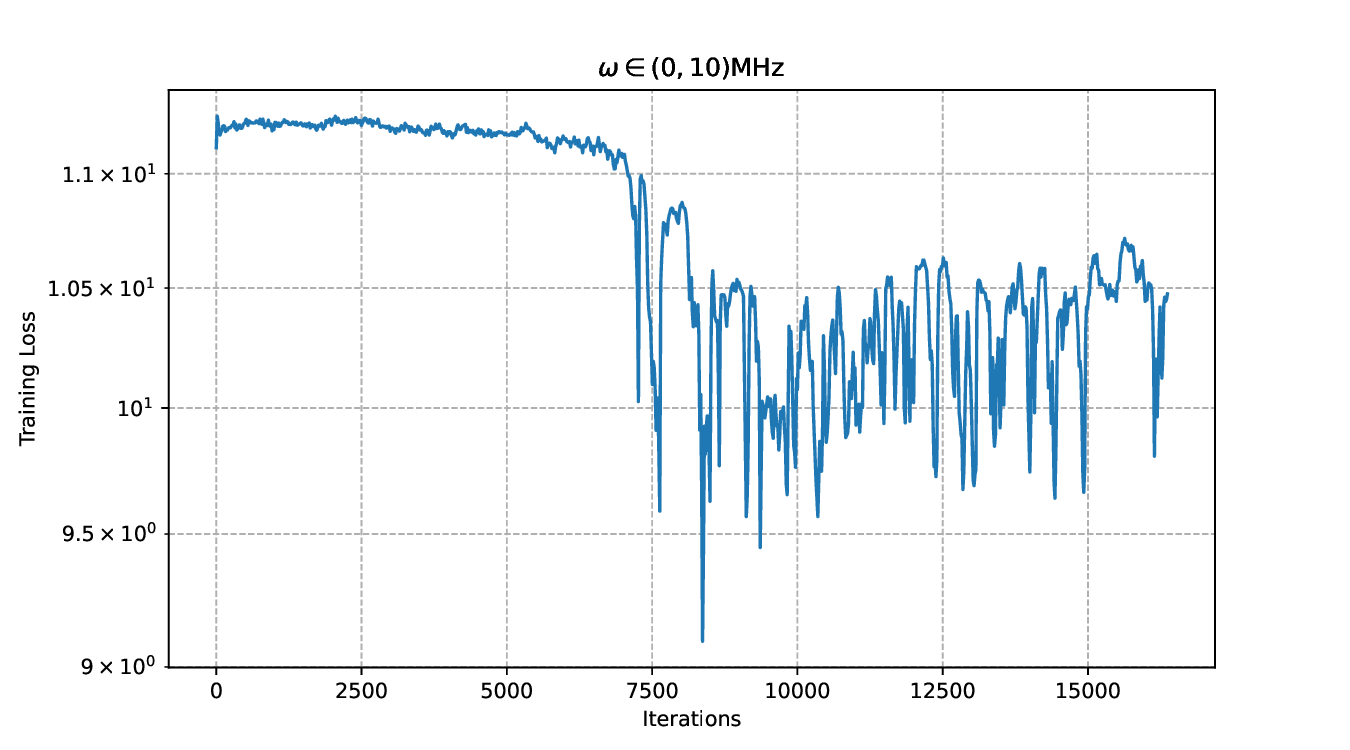}
		\caption*{Work by Belliardo, F., et al. \cite{belliardo2024model}}
	\end{subfigure}
	\begin{subfigure}[b]{0.31\textwidth}
		\includegraphics[width=\linewidth]{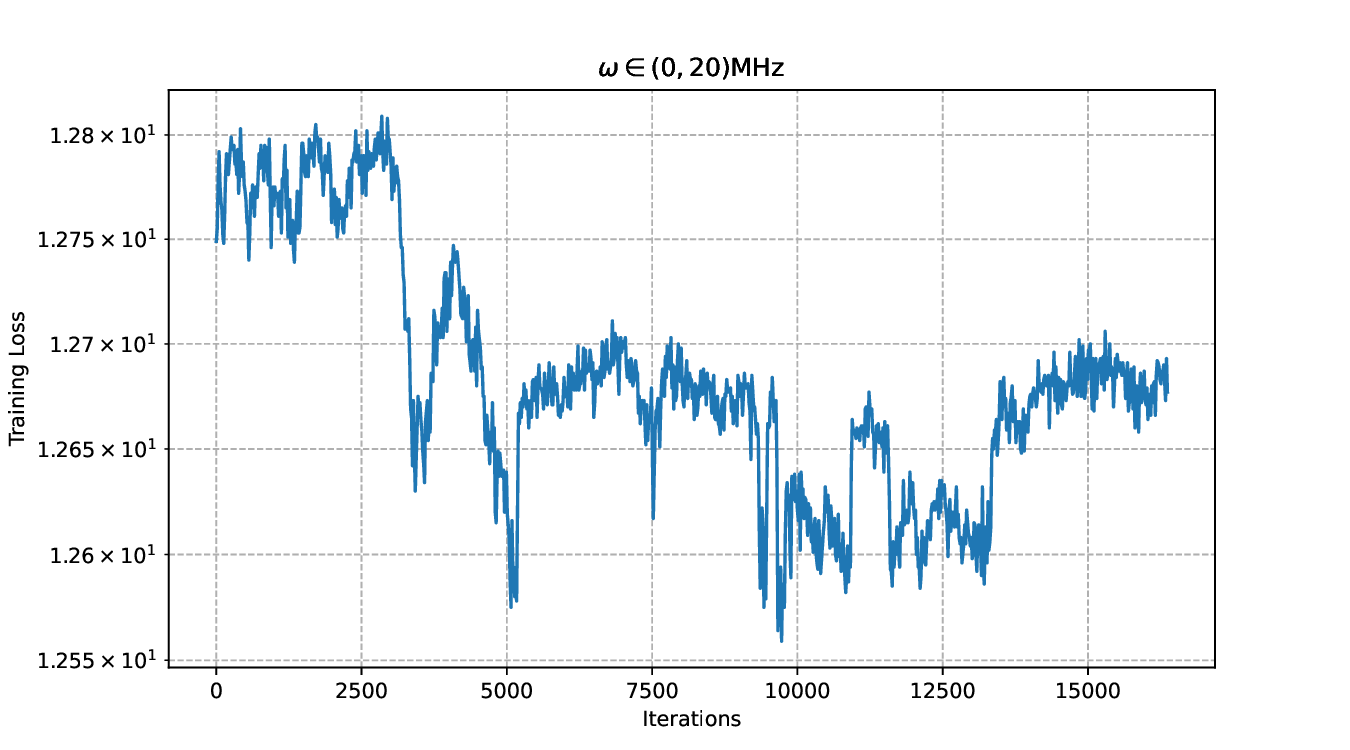}
		\caption*{}
	\end{subfigure}
	\vspace{5pt} 
	\begin{subfigure}[b]{0.31\textwidth}
		\includegraphics[width=\linewidth]{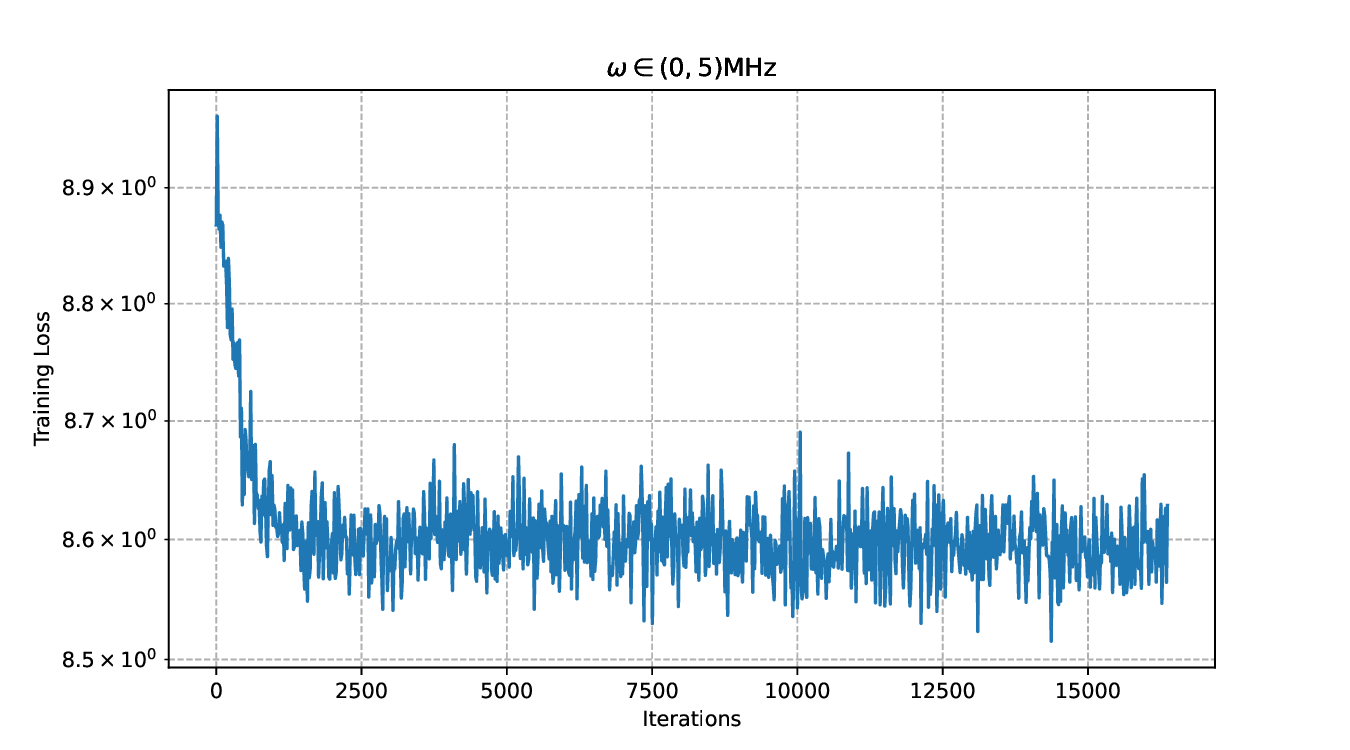}
		\caption*{}
	\end{subfigure}
	\begin{subfigure}[b]{0.31\textwidth}
		\includegraphics[width=\linewidth]{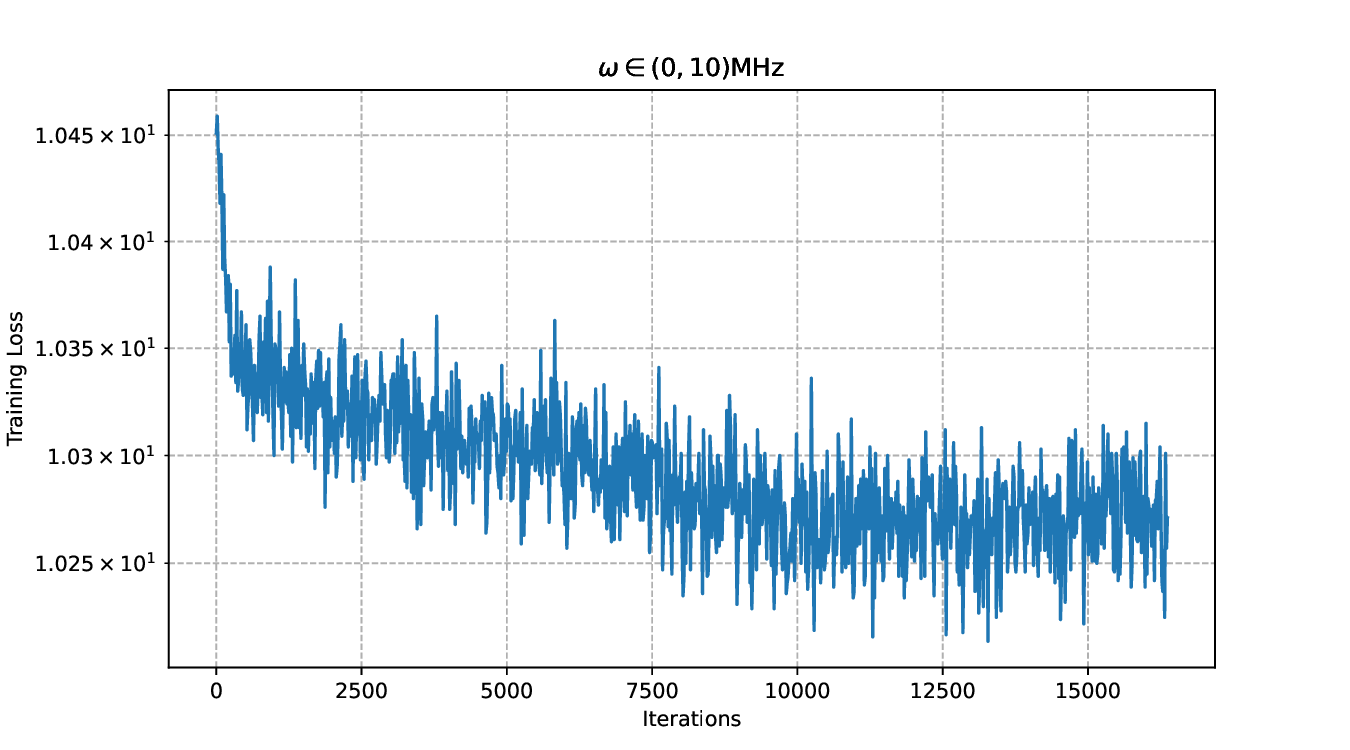}
		\caption*{Work by Bonato, C., et al. \cite{bonato2016optimized}}
	\end{subfigure}
	\begin{subfigure}[b]{0.31\textwidth}
		\includegraphics[width=\linewidth]{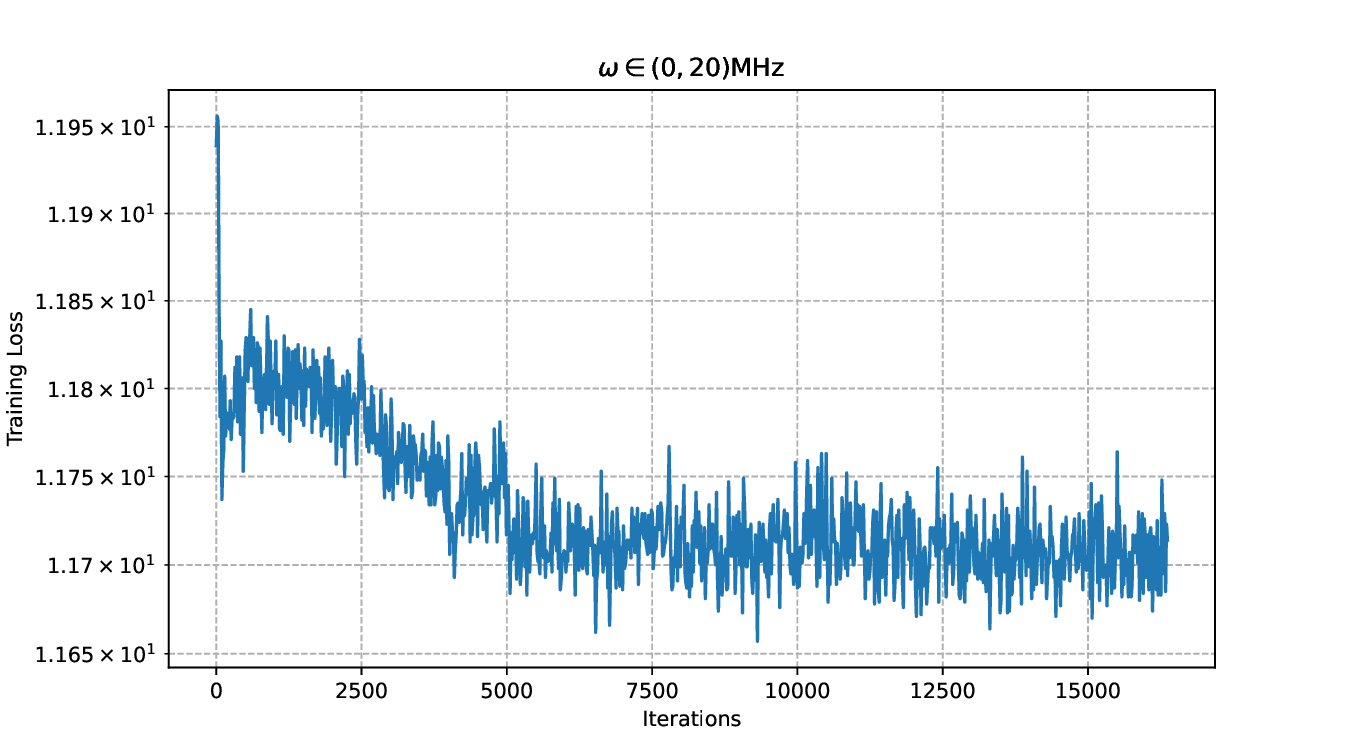}
		\caption*{}
	\end{subfigure}
	\captionsetup{name=Supplementary Figure}
	\caption{Training loss of the RL agent in the two baseline protocols}
	\label{fig:RL:loss}
\end{figure}

To further compare the performance of the two adaptive baseline methods by Belliardo et al.~\cite{belliardo2024model} and Bonato et al.~\cite{bonato2016optimized}, we examine the performance of their models in terms of training loss measured by the estimation MSE. Note that the swarm optimization method by Bonato et al.~\cite{bonato2016optimized} initially used Holevo variance to dictate iterative search of the sensing parameter $\varphi$. To make a fair comparison, we re-implement Bonato's method using the same RL framework as the one in Belliardo et al.~\cite{belliardo2024model}. The changes include but are not limited to using particle filter to approximate the Bayesian posterior probability and using MSE as the loss function. Some exemplary simulation setups are as follows: the number of particles was fixed at $P = 960$, and the number of training iterations is set to 16,384. Supplementary Figure \ref{fig:RL:loss} plots the training loss curves of the two methods. This result can explain the performance of the two methods in Figure 2 and 3 of the main document. Because Bonato's method only optimizes $\varphi$, its model convergence is much better than Belliardo's method, particularly when $\omega \in (0,20)$MHz for wide-range sensing. Their training losses in three $\omega$ settings are rather close, with Belliardo's method having a minor edge for small $\omega$'s. 

\section{Optimization of $\Delta$}\label{secResult-delta}
The following simulation results to illustrate how performance is affected by varying \(\Delta\). 

	\begin{figure}[H]
		\centering
		\begin{subfigure}[t]{0.49\textwidth}
			\centering
			\includegraphics[width=\linewidth]{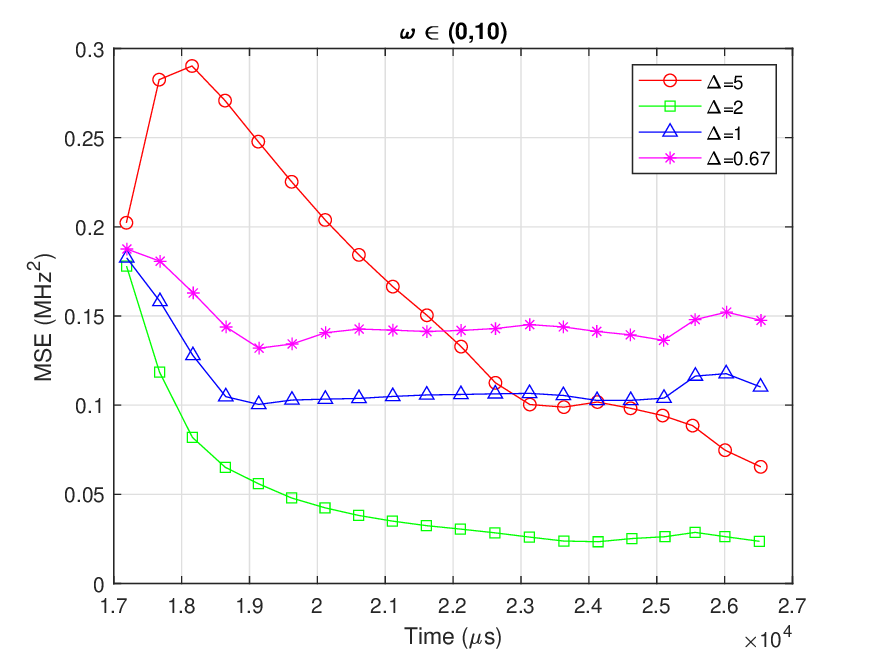}
			\caption{\(\Delta < \omega_{\max}\)}
			\label{fig:6a}
		\end{subfigure}
		\hfill
		\begin{subfigure}[t]{0.49\textwidth}
			\centering
			\includegraphics[width=\linewidth]{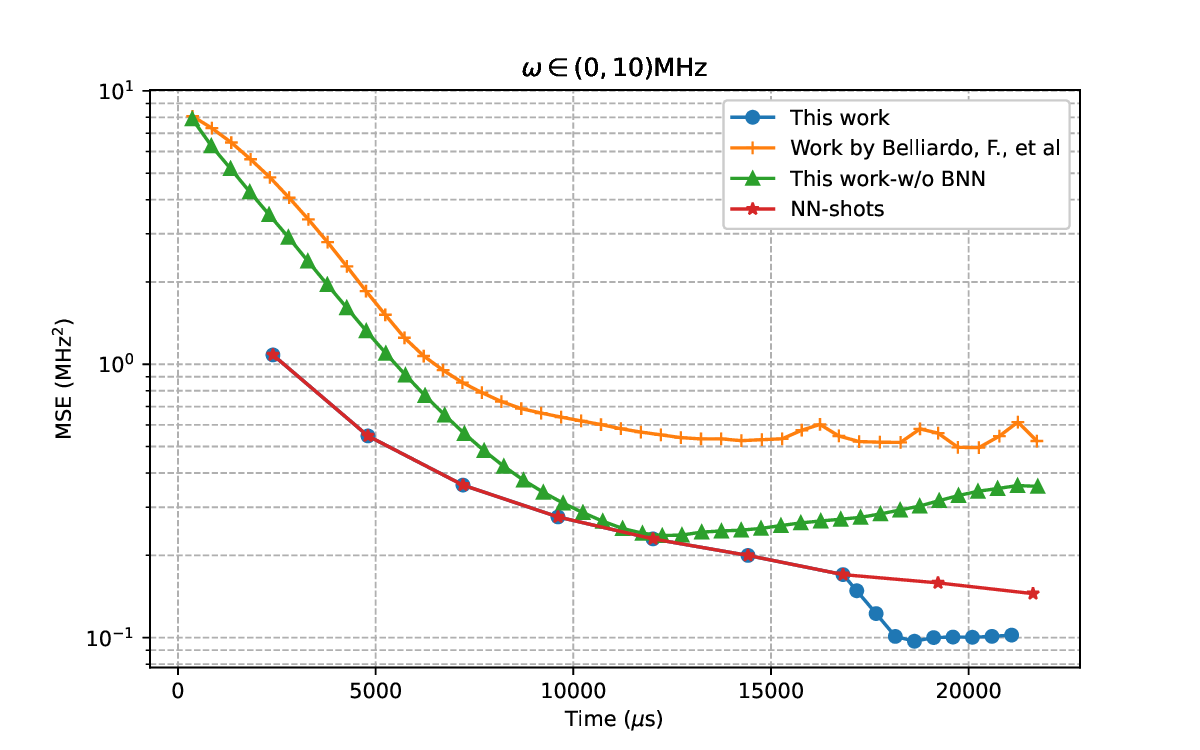}
			\caption{\(\Delta = \omega_{\max}\)}
			\label{fig:6b}
		\end{subfigure}
        \captionsetup{name=Supplementary Figure}
		\caption{MSE performance with respect to different choices of \(\Delta\).}
		\label{fig:6}
	\end{figure}

    Supplementary Figure~\ref{fig:6} illustrates the sensitivity of performance with respect to the choice of \(\Delta\). To isolate the effect of \(\Delta\) and avoid confounding due to changes in particle resolution (\(\Delta/P\)), we scale the number of particles \(P\) proportionally with \(\Delta\): for \(\Delta = 5\), we set \(P = 240 \times 4\); for \(\Delta = 2\), \(P = 240 \times 2\); and for \(\Delta = 1\) and \(\Delta = 0.67\), \(P = 240\). This ensures a consistent particle density across the tested ranges, allowing a fair comparison of the impact of \(\Delta\) on estimation performance.

	Supplementary Figure~\ref{fig:6}(a) presents results for the field range \(\omega \in (0, 10)\,\mathrm{MHz}\), with a total time budget of \(22{,}000\,\mu\mathrm{s}\) and a non-adaptive first stage comprising 70 Ramsey measurements. The BNN estimation error is approximately \(\varepsilon = 0.41\,\mathrm{MHz}\). The case with \(\Delta = 0.67\,\mathrm{MHz}\) (purple curve) violates the lower bound \(2\varepsilon\), and its performance degrades significantly. In contrast, \(\Delta = 1\) and \(2\,\mathrm{MHz}\) satisfy the criterion and lead to better results. A larger \(\Delta\) increases the likelihood that the true \(\omega\) lies within the adaptive range, improving the effectiveness of both Bayesian updates and RL exploration. Supplementary Figure~\ref{fig:6}(b) illustrates performance when \(\Delta = \omega_{\max}\), effectively bypassing the first-stage BNN and relying solely on RL-based inference (“This work w/o BNN”). In this regime, large \(\Delta\) values (e.g., \(5\) and \(10\,\mathrm{MHz}\)) yield broad posterior distributions and inefficient exploration, resulting in slower convergence and reduced estimation precision. 
    
    This sensitivity analysis confirms the critical role of \(\Delta\): if too small (violating \(2\varepsilon \leq \Delta\)), it risks excluding the true parameter; if too large (approaching \(\omega_{\max}\)), it dilutes the efficiency of exploration. These findings validate our empirical choices and underscore the trade-off: \(\Delta\) must be wide enough to include the true \(\omega\) with high probability, yet narrow enough to guide focused and efficient learning.

\bibliography{sn-bibliography}

\end{document}